\documentclass[3p,review,pdftex]{elsarticle}
\usepackage{lineno,hyperref}

\journal{Additive Manufacturing}

\usepackage{amsmath}
\usepackage{amsthm}
\usepackage{amssymb}
\usepackage{eucal}
\usepackage{mathrsfs}
\usepackage{subfigure}

\usepackage{fix-cm}
\usepackage{bm}
\usepackage[dvipdfmx]{color}
\usepackage{amsfonts}
\usepackage{enumerate}
\usepackage{color}

\def\dO{\ \text{d}\hspace{-0.3mm} \Omega}
\def\dG{\ \text{d}\hspace{-0.3mm}\Gamma}

\def \div{\mbox{\rm div}}
\def\x{\bm{x}}

\def \DD {\mathbb D}

\def \RR {\mathbb R}

\def \H {\mathscr{H}}

\def \P {\mathscr{P}}
\def \R {\mathscr{R}}

\def \U {\mathscr{U}}

\begin{document}

\begin{frontmatter}

\title{Topology optimization with a closed cavity exclusion constraint for additive manufacturing based on the fictitious physical model approach}

\author[mymainaddress]{T. Yamada \corref{mycorrespondingauthor}}
\ead{t.yamada@mech.t.u-tokyo.ac.jp}
\author[mymainaddress]{Y. Noguchi}

\cortext[mycorrespondingauthor]{Corresponding author}
\address[mymainaddress]{Department of Strategic Studies, Institute of Engineering Innovation, Graduate School of Engineering, The University of Tokyo, 2-11-16 Yayoi, Bunkyo-ku, Tokyo 113-8656, Japan.}

\begin{abstract}
This paper proposes a topology optimization method that considers the geometric constraint of no closed cavities to improve the effectiveness of additive manufacturing based on the fictitious physical model approach. First, the basic topology optimization concept and level set-based method are introduced. Next, the fictitious physical model for a geometric constraint in the topology optimization framework is discussed. Then, a model for the geometric constraint of no closed cavities for additive manufacturing is proposed. Numerical examples are provided to validate the proposed model. In addition, topology optimization considering the geometric constraint is formulated, and topology optimization algorithms are constructed using the finite element method. Finally, optimization examples are provided to validate the proposed topology optimization method.
\end{abstract}

\begin{keyword}
Topology optimization\sep Additive manufacturing\sep Hollow exclusion constraint\sep Partial differential equations\sep Finite element method\sep FreeFEM\sep Level set method
\end{keyword}

\end{frontmatter}

\section{Introduction}
This paper proposes a topology optimization method considering a geometric constraint for additive manufacturing based on a fictitious physical model. Additive manufacturing has attracted attention as a manufacturing process that has a higher degree of freedom than standard milling. The design phase is undergoing fundamental revision as design proposals that were previously difficult to manufacture can be realized. The design method for additive manufacturing is called Design for Additive Manufacturing (DfAM), and topology optimization is a powerful tool for improving this method.

Topology optimization \cite{bendsoe1988generating,suzuki1991homogenization} is a method to create and design a shape that maximizes or minimizes a given evaluation function without relying on a trial-and-error approach based on the designer’s experience or mechanical considerations. Compared with size optimization, which optimizes the given shape parameters, and shape optimization, which determines the optimal shape by changing the shape boundaries of a given shape, topology optimization has the highest degree of design freedom, allowing the search for the optimal design solution while also allowing changes in the topology, such as the number of shape boundaries. However, topology optimization results in geometrically complex shapes that are difficult to manufacture. To overcome this problem, the combination of topology optimization and additive manufacturing has recently attracted significant attention \cite{mezzadri2018topology,miki2021topology,wu2017minimum}.

The problem that has received the most attention is the overhang shape constraint. In general additive manufacturing, the shape of an object is considered to be a two-dimensional shape divided into equal intervals by planes parallel to the horizontal plane, and the three-dimensional shape of the object is formed by stacking the layers of the two-dimensional shape from the bottom up. Due to the characteristics of this manufacturing process, it is difficult to stack material on top of the void domain. When manufacturing a shape that protrudes into the void domain, called an overhang shape, it is necessary to build a temporary structure to support the weight of the object, called a support material. The support material must be removed by machining at the end of the manufacturing process, which not only increases the manufacturing time and cost, but also makes it difficult to remove the support material depending on the object shape.

Several methods of incorporating geometric constraints into topology optimization that eliminate overhang shapes have been investigated. Leary et al.  \cite{leary2014optimal} proposed a method of modifying the shape obtained by topology optimization to obtain a design solution that satisfies the overhang constraint. Langelaar \cite{langelaar2016topology} and Gaynor et al. \cite{gaynor2016topology} proposed a method of restricting the design space so that the shape is constructed only within the allowed overhang angle during the optimization process. Allaire et al. \cite{allaire2017structural} and Wang et al. \cite{wang2018level} reported that the overhang constraint cannot be satisfied only by constraining the normal direction of the model geometry.

Of metal additive manufacturing methods, laser powder bed fusion (LPBF) \cite{king2014observation,khairallah2016laser} is the most popular. In LPBF, a laser beam is irradiated into a container filled with metallic powder to locally melt and solidify the metallic powder to form the desired shape. The remaining metal powder must then be removed; however, for hollow shapes, such as eggshells, the metal powder cannot be removed. Although it is possible to create a small hole to remove the metal powder, this increases the number of processes and reduces the performance. Specifically, the process is extremely complex when hollow shapes overlap many times. Therefore, it is necessary to eliminate hollow shapes in the shape design stage while maintaining high performance.
To overcome this challenge, Liu et al. \cite{liu2015identification} proposed a virtual temperature method to obtain a simple connected structure. Although this method produces a single connected shape, it has the problem of imposing excessive shape constraints because the distribution of the temperature field depends on the size and aspect ratio of the domain. Li et al. \cite{li2016structural} proposed a virtual scalar field method; however, this approach is almost identical to the method of Liu et al. and has the same drawbacks. Zhou et al. \cite{zhou2019topology} proposed a method to provide the desired geometric constraints by imposing side constraints on the design variables. Although the constraints are not nonlinear, they have the problem of leading to an overly localized solution search. Unlike the direct geometric constraints of the method proposed by Zhou et al., virtual approaches are less prone to excessive locality in the solution search because the initial shape and intermediate solutions do not need to satisfy constraints. In addition, by setting an appropriate field, one can search for the shape without considering violating the constraints.

Virtual approaches are similar in concept to the fictitious physical model concept for geometric constraints in topology optimization \cite{sato2017manufacturability,yamada2018thickness,yamada2019geometric}.
The fictitious physical model is considered additional physics in topology optimization formulation. The fictitious physical model is formulated to represent the target geometric constraints; therefore, the target geometric constraints are evaluated in the same manner as normal physical properties. In other words, the target geometric constraints can be formulated in the standard topology optimization framework using the fictitious physical model.

This paper proposes an appropriate fictitious physical model of a no-closed-cavity constraint for additive manufacturing and a topology optimization method in consideration of the geometric constraint using a level-set-based method. The proposed fictitious field does not impose excessive constraints; however, it can be evaluated to be close to a constraint violation.

The remainder of this paper is organized as follows. First, we briefly discuss the level-set-based topology optimization method. Second, the fictitious physical model of the no-closed-cavity constraint is formulated. Here, numerical examples are presented to demonstrate the validity of the proposed model using the finite element method (FEM). Then, a topology optimization problem with the geometric constraint is formulated, and an optimization algorithm is constructed that uses the FEM when solving the governing equation, the fictitious physical model, and its adjoint problem. Finally, three-dimensional numerical examples are presented to demonstrate the utility of the proposed method.

\section{Level set-based topology optimization}
Here, we briefly discuss the level-set-based topology optimization method \cite{yamada2010topology}. The basic concept of topology optimization is the replacement of a structural optimization problem by a material distribution problem \cite{bendsoe1988generating}. Here, we define domain $D$, in which the designed structure can be placed. The domain $D\subset \RR^d$ is referred to as the fixed design domain because it is not changed in the process of finding the design solution. The target design shape $\Omega_1$ in the fixed design domain $D$ is distinguished using the following characteristic function $\chi \in L^{\infty}(D)$:
\begin{equation}
	\chi(\x):=\begin{cases}1\qquad&{\rm for}\quad \x \in \Omega_1\\
		0 \qquad&{\rm for}\quad \x \in D\setminus \Omega_1
	\end{cases}
\end{equation}
Although any shape and topological change can be represented by changing the characteristic function, an infinitely small structure is also permitted. Therefore, the topology optimization problem is an ill-posed problem \cite{allaire2012shape}.
The homogenized design method \cite{bendsoe1988generating,suzuki1991homogenization} and density based-method \cite{bendsoe1999material} are popular techniques to overcome this problem using relaxation and regularization methods, respectively.

In addition, level set-based structural optimization methods \cite{sethian2000structural,wang2003level,allaire2004structural,luo2008level,wang2018velocity} are also popular. In these methods, the structural shape is represented by the iso-surface of a distributed scalar function, that is, the so-called level set function. The level set function is updated using the extended shape sensitivity in level set-based shape optimization methods \cite{wang2003level,allaire2004structural}.
The current study employs level set-based topology optimization using topological derivatives \cite{yamada2010topology,amstutz2006new,allaire2005structural}. In particular, we select the method based on a reaction diffusion equation \cite{yamada2010topology}.
The level set function is defined as follows:
\begin{equation}
	\begin{cases}
		0< \phi(\x) \le 1\qquad &{\rm for}\quad \x \in \Omega_1 \setminus \partial \Omega_1\\
		\phi(\x)=0\qquad &{\rm for}\quad \x \in \partial \Omega_1\\
		-1\le \phi(\x) <0 \qquad & {\rm for} \quad \x \in D\setminus \Omega_1
	\end{cases}
\end{equation}
The level set function $\phi$ is a scalar function that takes the positive and negative values in the material domain $\Omega_1$ and the void domain, respectively. The boundary between the material and void domains is represented by the zero iso-surface of the level set function.  
The characteristic function is redefined using the level set function $\phi$.
\begin{equation}
	\chi_\phi:=
	\begin{cases}
		1\qquad &{\rm for} \quad \phi(\x)\ge 0\\
		0\qquad &{\rm for} \quad \phi(\x)< 0
	\end{cases}
\end{equation}
The topology optimization problem is formulated as an optimization problem in which the design variable is the level set function.
\\
Next, the method used for optimizing level-set-based topology optimization \cite{yamada2010topology} is briefly discussed. It is difficult to directly determine the distribution of the level-set function representing an optimal shape. Therefore, to find the optimal solution, the problem of determining the distribution of an optimal level-set function is replaced by solving the time evolution equation. Fictitious time $t$ is introduced, and the level-set function is updated based on the following reaction diffusion equation to find the optimal level-set function:
\begin{equation}
	\frac{\partial \phi }{\partial t}=-K\left( J'-\tau \nabla^2 \phi \right)
\end{equation}
where $K>0$ is a proportional constant, $J'$ is the topological derivative of the target optimization problem, and $\tau>0$ is a regularization parameter. Note that the geometric complexity of the obtained optimal configuration can be controlled by adjusting regularization parameter $\tau$.

In addition, an appropriate boundary condition is considered, and we obtain the following:
\begin{equation}
	\begin{cases}
		\frac{\partial \phi }{\partial t}=-K\left( J'-\tau \nabla^2 \phi \right)
		\qquad&{\rm in}\quad D\\
		\nabla \phi \cdot \bm{n}=0\qquad &{\rm on}\quad \partial D\setminus \partial D_m\\
		\phi=1\qquad&{\rm on}\quad \partial D_m
	\end{cases}
\end{equation}
where $\bm{n}$ is the unit normal vector on boundary $\partial D$, and part of boundary $\partial D_m$ is connected to the material domain outside of the fixed design domain $D$.
We obtain the optimal solutions by solving the above equation in each optimization step.
Based on the reaction-diffusion equation, the gradient method is used for optimization. In addition, a small diffusion term is added for regularization. The diffusion term smoothens the shape and controls the geometric complexity of the obtained shape by adjusting its magnitude.
\section{Geometric constraint of no closed cavities}
\subsection{Geometric constraint based on the fictitious physical model}
First, we briefly discuss the basic topology optimization framework. The framework is an optimization problem with the governing equation giving the state variable $u$ as the constraint, as demonstrated in the following equation:
\begin{align}
	\min\qquad&J[u]\notag\\
	\textrm{subject to:}\qquad&\textrm{governing equation for $u$},\notag
\end{align}
where $J$ is an objective function. For example, the elasticity equation is the constraint equation where the objective of the optimization problem is to minimize the maximum stress, and the Navier Stokes equation is the constraint equation where the objective of the design is the lift or drag force of the target wing shape. In other words, the basic framework of topology optimization is that the evaluation indices, such as the objective function and constraint function, are formulated by state variables. Furthermore, it is assumed that numerical analysis of the governing equations that give the state variables is possible. 
As a result of this framework, it is difficult to directly consider manufacturability and assemblability, for which state variables do not exist explicitly, although topology optimization can obtain optimal design solutions from a mechanical perspective. To overcome this problem, the fictitious physical model has been proposed \cite{yamada2018thickness, sato2017manufacturability}. In this approach, a fictitious physical field and its governing equations are considered to evaluate manufacturability. That is, the basic ideas are the (i) introduction of the fictitious physical field and (ii) formulation of an evaluation function for the target geometric constraint using the field variable. In addition, an optimization problem with the geometric constraint is formulated as a multiphysics problem comprising both standard and fictitious physics. This concept allows for optimal design for manufacturability in the basic framework of topology optimization.
%
\subsection{Partial differential equation (PDE) for target geometric constraint}
\begin{figure}[htb]
	\begin{center}
		\includegraphics[height=4cm]{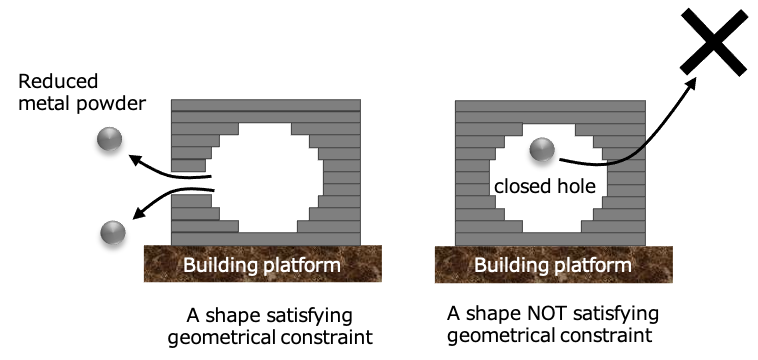}
		\caption{Conceptual shapes for the geometric constraint.}
		\label{fig:concept}
	\end{center}
\end{figure}

The geometric constraint of a no closed holes is formulated using the fictitious physical model concept. 
The geometric constraint of no closed cavities is formulated using the fictitious physical model. In LPBF and other additive manufacturing processes, the target shape is built on a building platform layer by layer. In LPBF, the process is to sinter the prelaid metal powder by irradiating it with a laser, and metal powder that is not sintered must be removed at the end. Therefore, any shape that cannot be removed from metal powder is not an effective design shape for LPBF.
For example, the shape illustrated on the left side in Figure \ref{fig:concept} can have the metal powder removed from the inside, whereas the shape illustrated on the right side cannot.\\
\\
Next, the fictitious physical variable $p \in H^{1}(D)$ and its governing equation are defined as follows:
\begin{align}
	\begin{cases}
		-\div \left(a_p  \nabla p \right)+\left(1-\chi\right)\left(p-1\right)=0\, \, \, \, & \text{in}\, \, \, \, D\\
		p=0\, \, \, \,& \text{on}\, \, \, \, \Gamma_p\\
		\nabla p \cdot \bm{n} =0\qquad&  \text{on}\quad \partial D \setminus \Gamma_p 
	\end{cases}
	\label{eq:to_Gvpw}
\end{align}
where
$$a_p :=\left\{\left(\bar{a}_{p}-\epsilon_{p} \right)\left(1-\chi\right)+\epsilon_{p}\right\}L^2$$
is a diffusion coefficient. $\bar{a}_{p}>0$ and $0<\epsilon_{p}<<1$ are diffusion coefficients in the void domain that take small values to represent small diffusion in the material domain.
In the later sections, we present numerical examples that confirm that $\epsilon_{p}$ should be set to a value of $\le 1\times 10^{-5}$. The characteristic length $L$ is set to the typical length of a substructure in the design object. In our numerical examples, the characteristic length $L$ is defined as the size of the fixed design domain.
The boundary $\Gamma_p$ indicates the boundaries connected to the outer side.

This PDE system represents fictitious steady state heat flow with a heat source in the void domain. This fictitious heat source is set so that the temperature is directed toward $1$ at each point in the void domain. In contrast, the thermal conductivity coefficient $\epsilon_{p}$ in the void domain is set to be large so that the heat diffuses immediately to the boundary $\Gamma_p$. Therefore, if the void domain is connected to the boundary $\Gamma_p$, the temperature is almost zero. Otherwise, the temperature approaches $1$, because the void domain enclosed by the material domain $\Omega_1$ is adiabatic. That is, in domains where the geometric constraint is violated, the temperature tends toward $1$, whereas in domains where the geometric constraint is satisfied, the temperature is $0$. If the temperature is almost $0$ in the whole domain,  then the constraint is satisfied.

It should be noted that since the virtual field takes $1$ or $0$ everywhere, the constraints and their design sensitivity are evaluated without being affected by the size and aspect ratio of the target geometry or void domain.
An important feature of the proposed model is that it is not affected by geometric information other than the target geometric features. Meanwhile, the methods that use the steady-state heat equation \cite{liu2015identification,li2016structural} are affected by the width and size of the shape, and the geometric features that are completely unrelated to the geometric constraints can affect the shape search. That is, it imposes an additional obstacle to the process of searching for the optimal shape because it gives a policy of changing the shape irrespective of the target geometric constraints.

\subsection{Constraint function and its adjoint equation}
If the geometric constraints are satisfied, the fictitious physical field $p$ has a value of almost zero in the whole domain. In contrast, if the geometric constraint is not satisfied, the value of the fictitious physical field $p$ approaches $1$ in the domain where the constraint is violated.
In other words, the geometric constraint is satisfied if the physical field $p$ is zero in the whole domain. Therefore, the constraint function for the geometric constraint is defined as 
\begin{equation}
	J_h:= \int_{D} \R(p) \dO,
\end{equation}
where $\R$ is the ramp function and is
\begin{equation}
\R(p):=
\begin{cases}
	p \qquad&{\rm if}\quad p\ge 0\\
	0 \qquad&{\rm if}\quad p < 0.
\end{cases}
\end{equation}
 We note that the definition of the constraint function is not unique. For example, the Heaviside function can be used instead of the ramp function. In this case, the derivative becomes the Kronecker delta function, which is not appropriate when using the gradient-based method.
Ramp functions are the standard choice \cite{wang2018level, wang2017structural} for formulating constraints in topology optimization.

The adjoint variable $\hat{p}\in H^{1}(D) $ is defined as follows.
\begin{align}
	&-\div \left(a_p  \nabla \hat{p} \right)+\left(1-\chi\right) \hat{p} =-\H(p) \qquad& & \text{in}\quad D
	\label{eq:adj-p}\\
	&\hat{p}=0\qquad& & \text{on}\quad \Gamma_p\\
	&\nabla \hat{p}\cdot \bm{n} =0\qquad& & \text{on}\quad \partial D \setminus \Gamma_p,
\end{align}
where $\H$ is the Heaviside unit function.
Then, the topological derivative \cite{novotny2003topological, carpio2008solving} relative to function $J_h$ is derived as follows:
\begin{equation}
	{J'_{h}=A_p(\chi) \nabla p \cdot \nabla \hat{p} - \left(p - 1\right) \hat{p},}
	\label{eq:td}
\end{equation}
where
\begin{eqnarray}
	{A_p}(\chi):=
	\begin{cases}
		\frac{2\bar{a}_{p}(\epsilon_p - \bar{a}_p)}{ \bar{a}_p + \epsilon_p  }L^2(1-\chi) 
		-\frac{2\epsilon_p(\bar{a}_p  - \epsilon_p)}{ \epsilon_p + \bar{a}_p }L^2\chi 
		\, \, &\textrm{if}\, \, d=2\\
		\frac{3\bar{a}_{p}(\epsilon_p - \bar{a}_p)}{ 2\bar{a}_p + \epsilon_p  }L^2(1-\chi) 
		-\frac{3\epsilon_p(\bar{a}_p  - \epsilon_p)}{ 2\epsilon_p + \bar{a}_p }L^2\chi 
		\, \, &\textrm{if}\, \, d=3.
	\end{cases}\notag
\end{eqnarray}
In this study, we multiply the second term of the topological derivative (\ref{eq:td}) by $(1-\chi)$ and use this as the approximate topological derivative for numerical stability.
%
\section{Formulation of topology optimization with geometric constraint}
\subsection{Minimum mean compliance problem}
First, the minimum mean compliance problem with the geometric and volume constraints is considered using the level set-based topology optimization method.
We assume that the material domain $\Omega_1$ is filled with a linear isotropic homogeneous material.
The displacement $\bm{u}$ is fixed at boundary $\Gamma_u$, and traction $\bm{t}$ is imposed at boundary $\Gamma_t$. The upper limit of the volume constraint is $V_{\max}$. Then, the optimization problem can be formulated as follows:
\begin{align}
	\inf_{\phi }\qquad
	&J_u=\displaystyle \int_{\Gamma_t} \bm{t}\cdot \bm{u} \dG \\
	\text{subject to: } &E_u=\int_{\Gamma_t} \bm{t} \cdot \tilde{\bm{u}} \dG
	-\int_D \bm{\epsilon}(\bm{u}): \DD \chi_\phi :\bm{\epsilon}(\tilde{\bm{u}}) \dO=0\notag\\
	&\qquad \text{for  }\, \, \forall \tilde{\bm{u}} \in \U, \bm{u} \in \U \notag\\
	&E_p=\int_D a_p \nabla p \cdot \nabla {\tilde{p} } \dO
	+\int_D (1-\chi_p)(p-1)  {\tilde{p} }\dO =0\notag\\
	&\qquad \text{for  }\, \,  \forall \tilde{p} \in \P, p \in \P \notag\\
	&J_h [p]=0\\
	&G_{vol }=\int_D \chi_{\phi} \dO -V_{\max} \le 0
\end{align}
where $\DD$ and $\bm{\epsilon}$ are the elastic tensor and linearized strain tensor, respectively. The functional spaces $\U$ and $\P$ are defined as follows.
\begin{align}
	&	\U= \left\{ \tilde{\bm{u}} \in H^1(D)^d \quad \Bigl |  \quad \tilde{\bm{u}}  =0\quad {\rm on }\quad \Gamma_u    \right\}\\
	&		\P= \left\{ \tilde{p} \in H^1(D) \quad \Bigl |  \quad \tilde{p}  =0\quad {\rm on }\quad \Gamma_p    \right\}
\end{align}

The topological derivative of the minimum mean compliance problem in three dimensions \cite{bonnet2013topological} is
\begin{equation}
J_u'=\bm{\epsilon}(\bm{u}): \mathbb{A} \chi_\phi :\bm{\epsilon}(\tilde{\bm{u}}),
\end{equation}
The coefficient tensor $\mathbb{A}$ is 
\begin{equation}
\mathbb{A}_{ijkl}=\frac{3(1-\nu)}{2(1+\nu)(7-5\nu)}\left[
-\frac{(1-14\nu+15\nu^2)E}{(1-2\nu)^2}\delta_{ij}\delta_{kl}
+5E(\delta_{ik}\delta_{jl}+\delta_{il}\delta_{jk})
\right],\notag
\end{equation}
where $E$, $\nu$ and $\delta_{ij}$ are the Young's modulus, Poisson ratio and Kronecker delta, respectively. We note that this optimization problem is a self-adjoint problem; that is, the adjoint equation is equivalent to the governing equation. The derivation details are provided in a previous paper \cite{otomori2015matlab}.
\subsection{Thermal diffusivity problem}
Here, we consider the steady-state internal heat generation problem  \cite{yamada2011level}. We suppose that the material domain $\Omega_1$ is filled with an arbitrary linear thermal conductor. The void domain is also occupied by a weak linear thermal conductor. The temperature $u_t(\x)=\overline{T}$ is imposed at the boundary $\Gamma_{temp}$, and the internal heat source $Q$ is set in design domain $D$. Then, the maximum thermal diffusivity problem with geometric and volume constraints is formulated as follows:
\begin{align}
	\inf_{\phi }\qquad
	&J_t=\displaystyle \int_{D} Q {u_t} \dO \\
	\text{subject to: } &E_t=\int_{D} Q \tilde{u_t} \dO
	-\int_D  \kappa \nabla u_t \cdot \nabla \tilde{u_t} \dO=0\notag\\
	&\qquad \text{for  }\, \, \forall \tilde{u_t} \in \U_t,\, \, u \in \U_t \notag\\
	&E_p=0\notag\\
	&\qquad \text{for  } \, \, \forall \tilde{p} \in \P,\, \, p \in \P \notag\\
	&J_h [p]=0\\
	&G_{vol } \le 0
\end{align}
where $\kappa$ is the linear thermal conduction tensor. The functional space $\U_t$ is defined as follows:
\begin{align}
	&	\U_t= \left\{ \tilde{u_t} \in H^1(D) \quad \Bigl |  \quad \tilde{u_t}  =0\quad {\rm on }\quad \Gamma_{temp}    \right\}
\end{align}
This problem is set to approximate the heat diffusion in a heat conductor with heat transfer boundary conditions, which is called thermal compliance \cite{bendsoe2003topology}.
The topological derivative of the thermal problem in three dimensions \cite{sokolowski1999topological} is
\begin{equation}
J_t'=\frac{2}{3}\kappa \nabla u_t \cdot \nabla u_t.
\end{equation}
We remark that this optimization problem is also a self-adjoint problem.

%
\section{Optimization algorithm}
The optimization algorithm of the proposed method is summarized as follows.
\begin{enumerate}[Step 1.]
	\item The level set function $\phi$ is initialized.
	\item The equilibrium equations for the target physics and fictitious field (\ref{eq:to_Gvpw}) are computed using the FEM.
	\item If the objective function converges, the optimization is completed. Otherwise, the adjoint equation of the fictitious physical model is computed using the FEM.
	\item The adjoint equation for the fictitious field (\ref{eq:adj-p}) is computed using the FEM.
	\item The design sensitivities are calculated.
	\item The level set function is updated by solving the reaction-diffusion equation using the FEM.
	The procedure returns to Step 2.
\end{enumerate}
Notably, the FreeFEM software \cite{hecht2012new} is used for all finite element analyses.
In the optimization examples presented in this study, the domain is discretized into tetrahedral elements. Because the physical and fictitious physical equations for geometric constraints are linear steady-state diffusion equations, no special numerical techniques are required. Therefore, numerical analysis using the FEM with hexahedral elements is also possible. In addition, the finite volume method or boundary element method can be applied for discretization.
%
\section{Numerical examples}
\subsection{Validation of proposed PDE}
\begin{figure}[htb]
	\begin{center}
		\includegraphics[height=3cm]{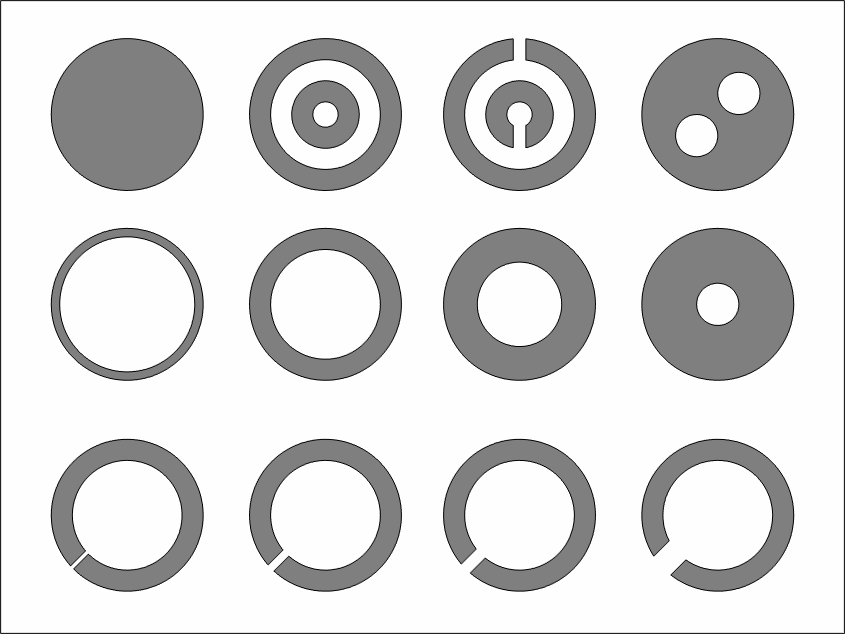}
		\caption{Material distribution for numerical example}
		\label{fig:chi}
	\end{center}
\end{figure}

To validate the proposed PDE for the geometric constraint of no closed cavities, numerical examples are provided using the FEM. Here, the material distribution displayed in Figure \ref{fig:chi} is considered.
The gray and white colors represent the material domain $\Omega_1$ and the void domain $D\setminus \Omega_1$, respectively. The characteristic length $L$ is set to $1$, because the size of the design domain $D$ has the dimensions of  $1.5\times2.0$. The design domain is discretized using a triangular mesh.

First, we examine the appropriate range and the effect of the parameter values with respect to the diffusion coefficient $a_p$, where the small diffusion coefficient $\epsilon_p$ is set to $1\times10^{-5}$.
\begin{figure*}[htb]
	\begin{center}
		\subfigure[$a_p=1\times10^{4}$]{
			\includegraphics[height=2.5cm]{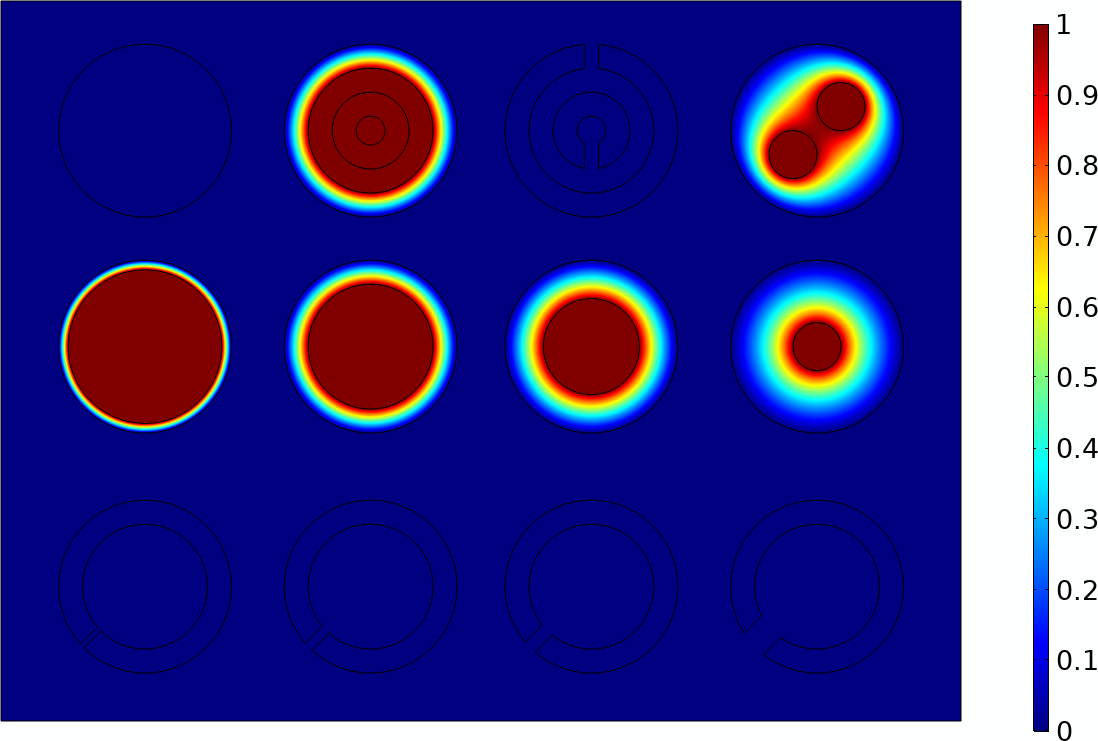}}
		\subfigure[$a_p=1\times10^{3}$]{
			\includegraphics[height=2.5cm]{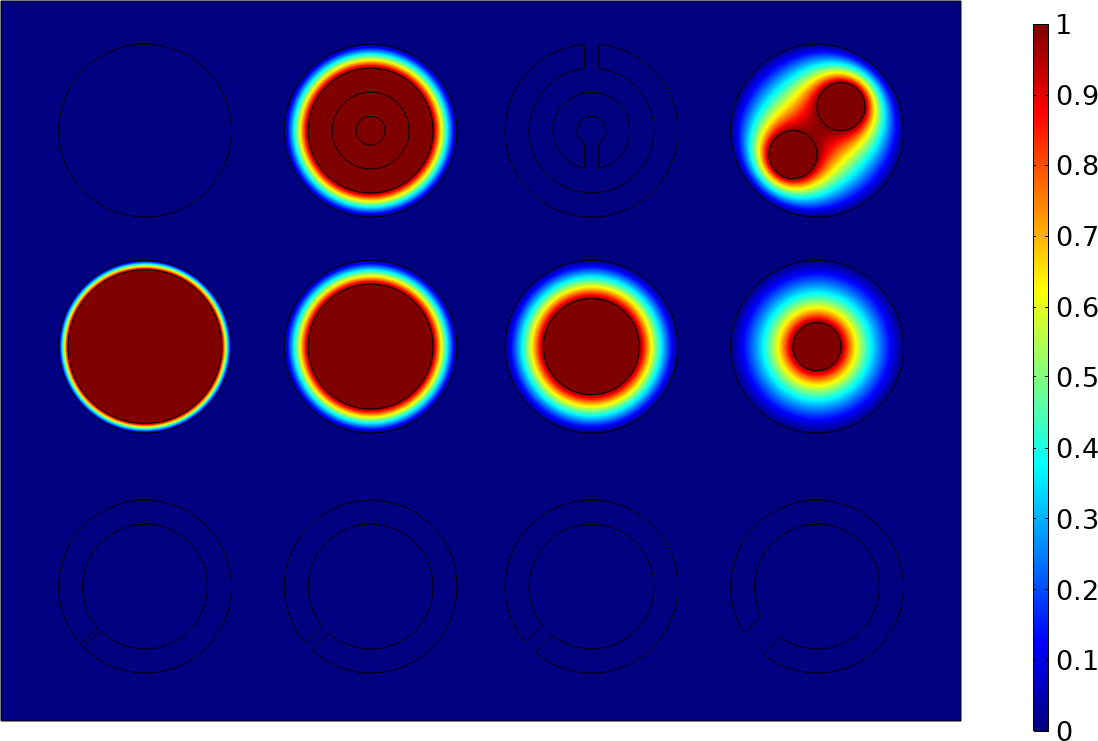}}
		\subfigure[$a_p=1\times10^{2}$]{
			\includegraphics[height=2.5cm]{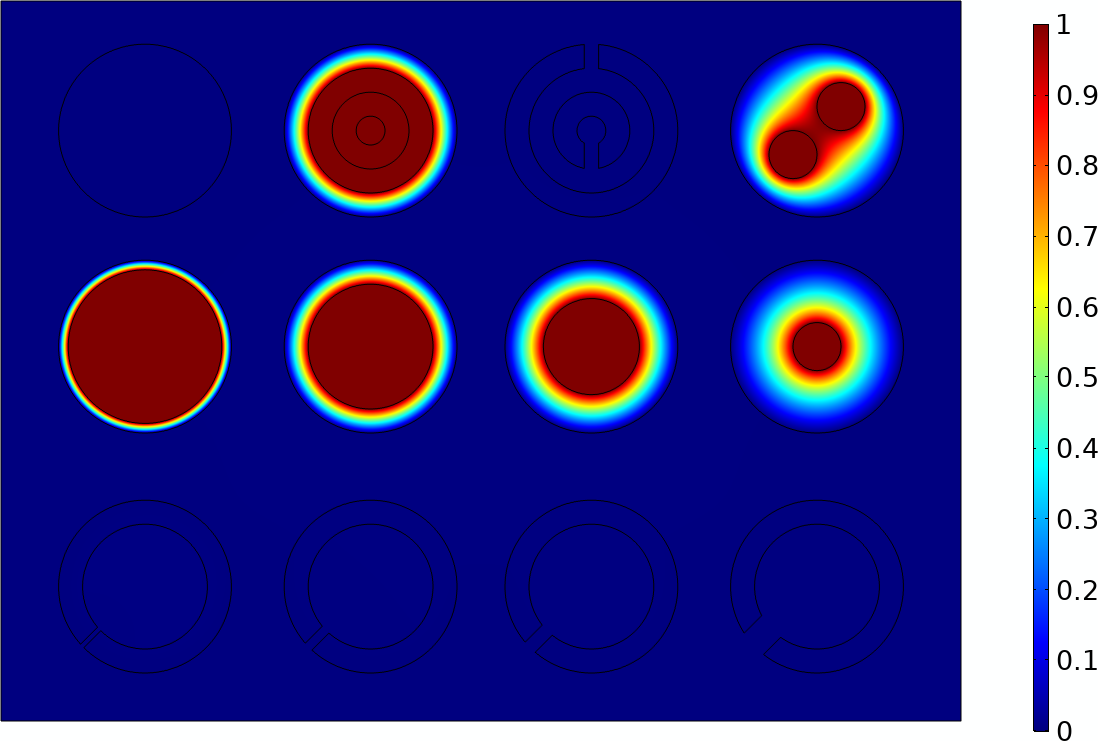}}
		\subfigure[$a_p=1\times10^{1}$]{
			\includegraphics[height=2.5cm]{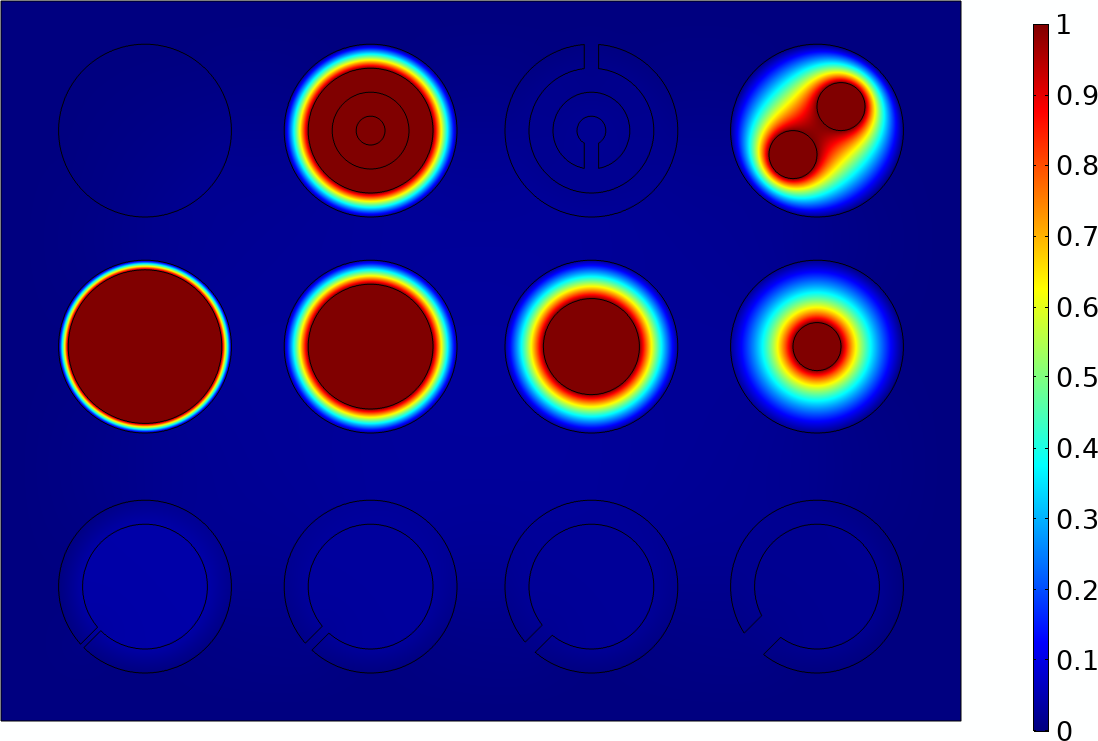}}
		\subfigure[$a_p=1\times10^{0}$]{
			\includegraphics[height=2.5cm]{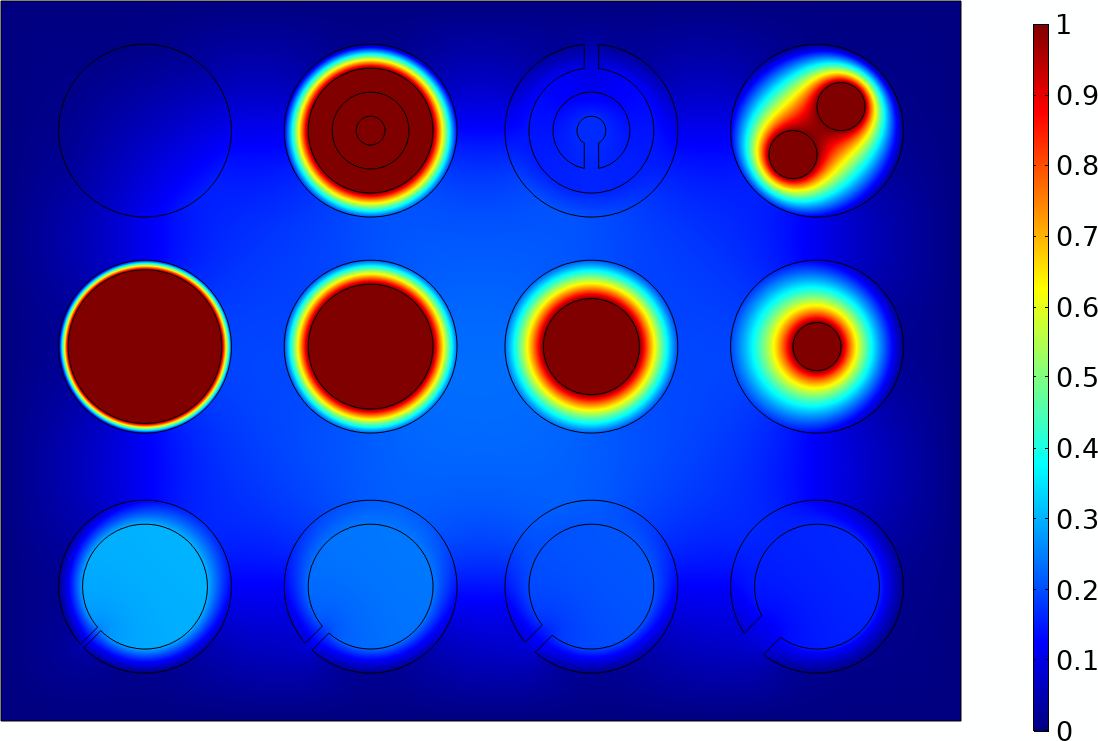}}
		\caption{Fictitious physical field $p$ for various values of the diffusion coefficient $a_p$}
		\label{fig:p-a}
	\end{center}
\end{figure*}
\begin{figure*}[htb]
	\begin{center}
		\subfigure[$a_p=1\times10^{4}$]{
			\includegraphics[height=2.5cm]{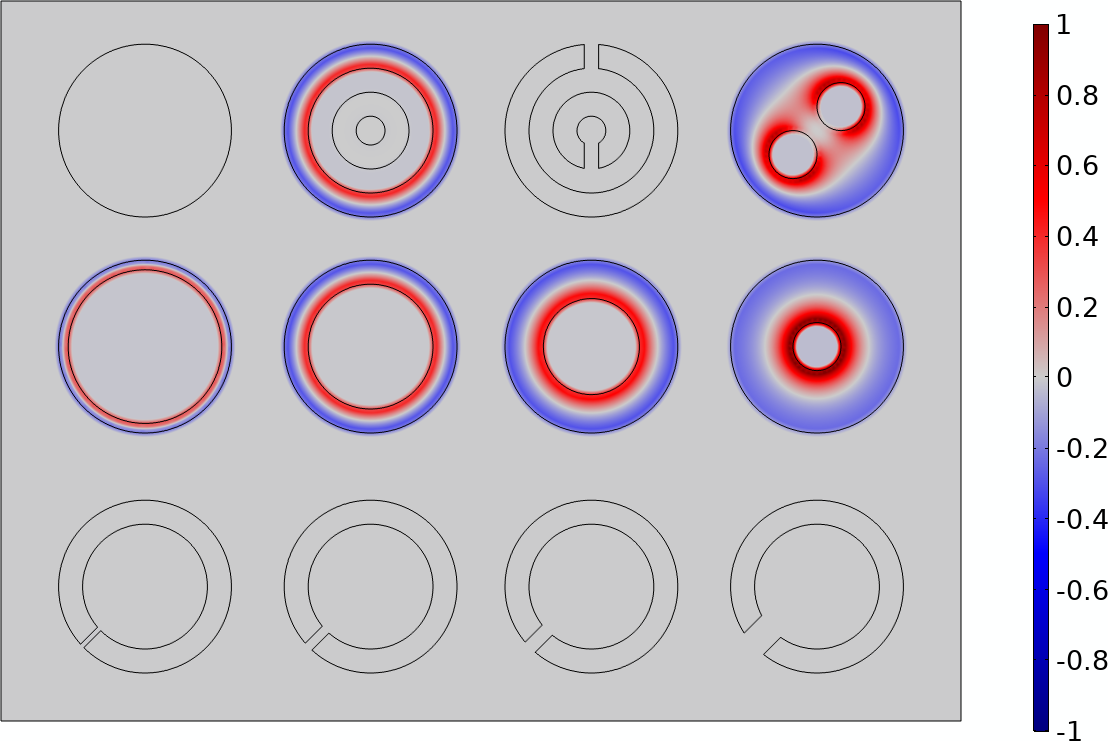}}
		\subfigure[$a_p=1\times10^{3}$]{
			\includegraphics[height=2.5cm]{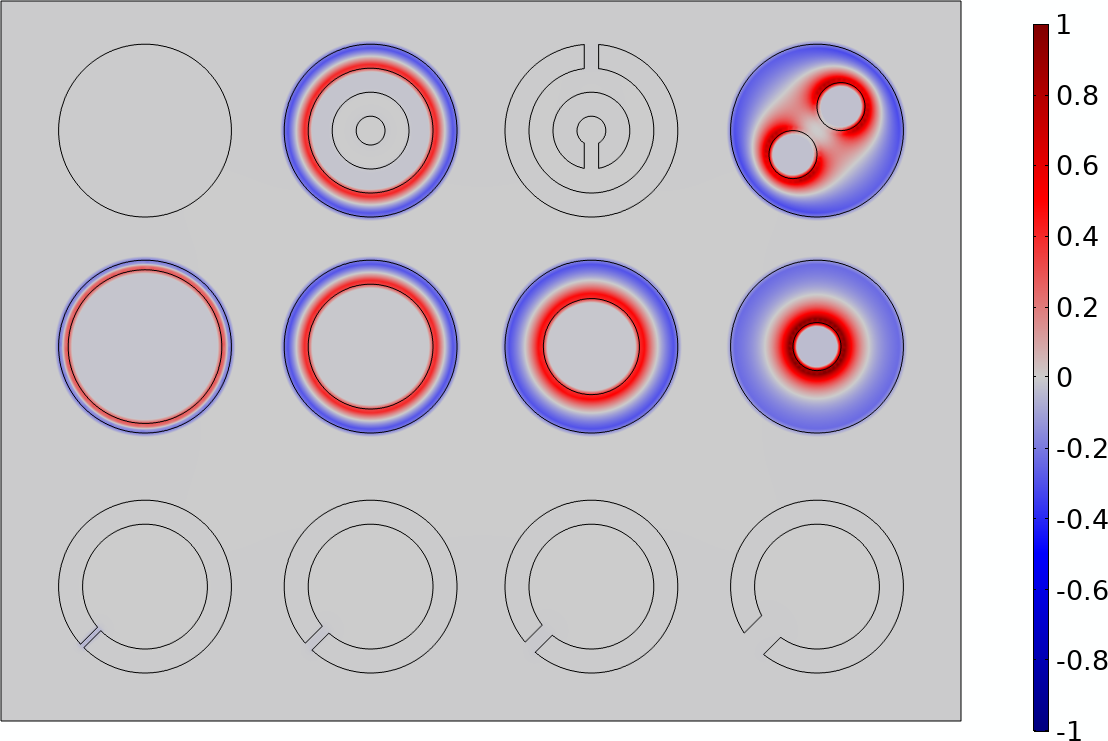}}
		\subfigure[$a_p=1\times10^{2}$]{
			\includegraphics[height=2.5cm]{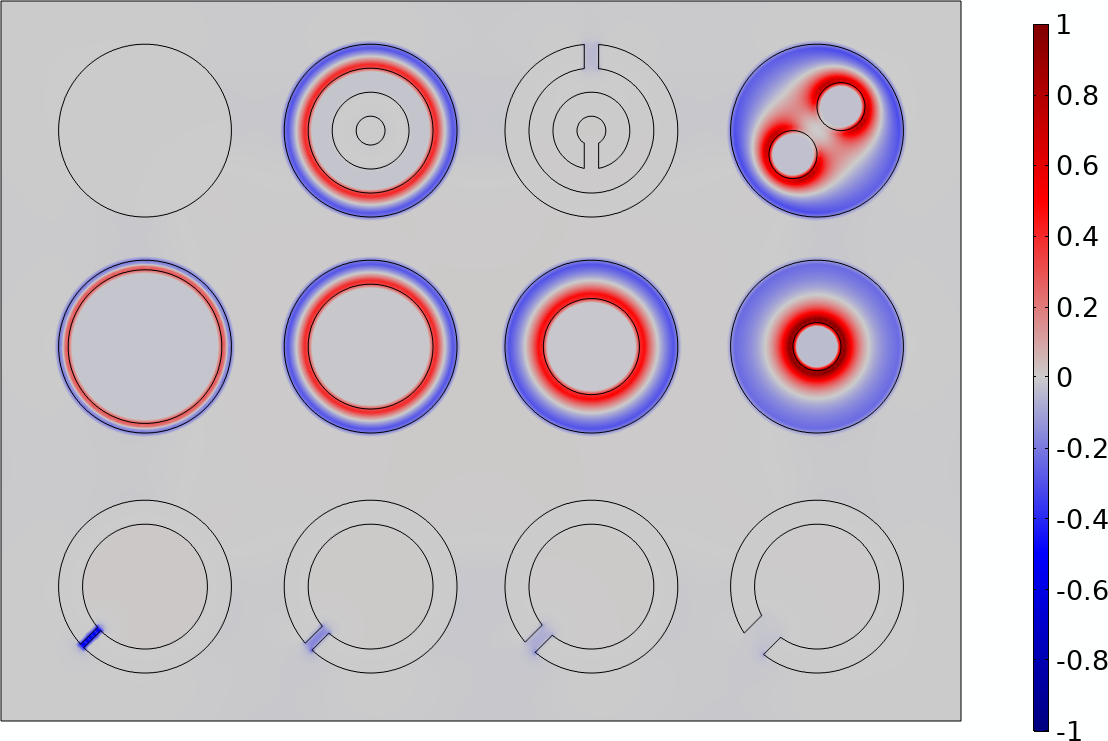}}
		\subfigure[$a_p=1\times10^{1}$]{
			\includegraphics[height=2.5cm]{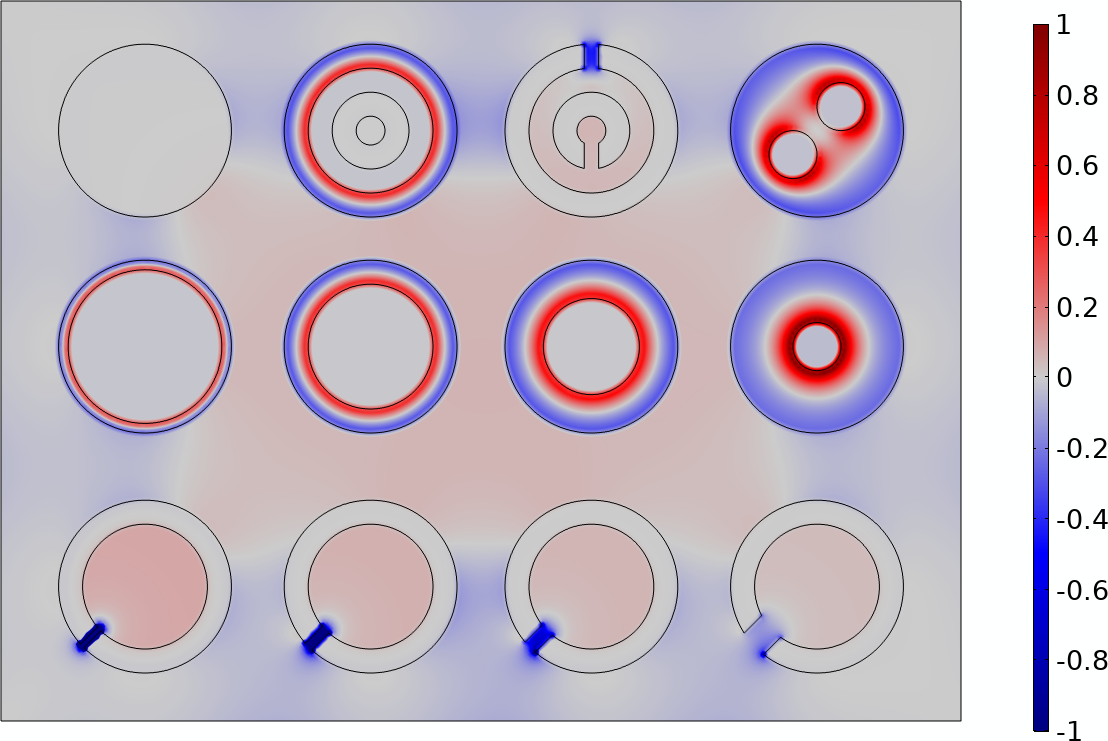}}
		\subfigure[$a_p=1\times10^{0}$]{
			\includegraphics[height=2.5cm]{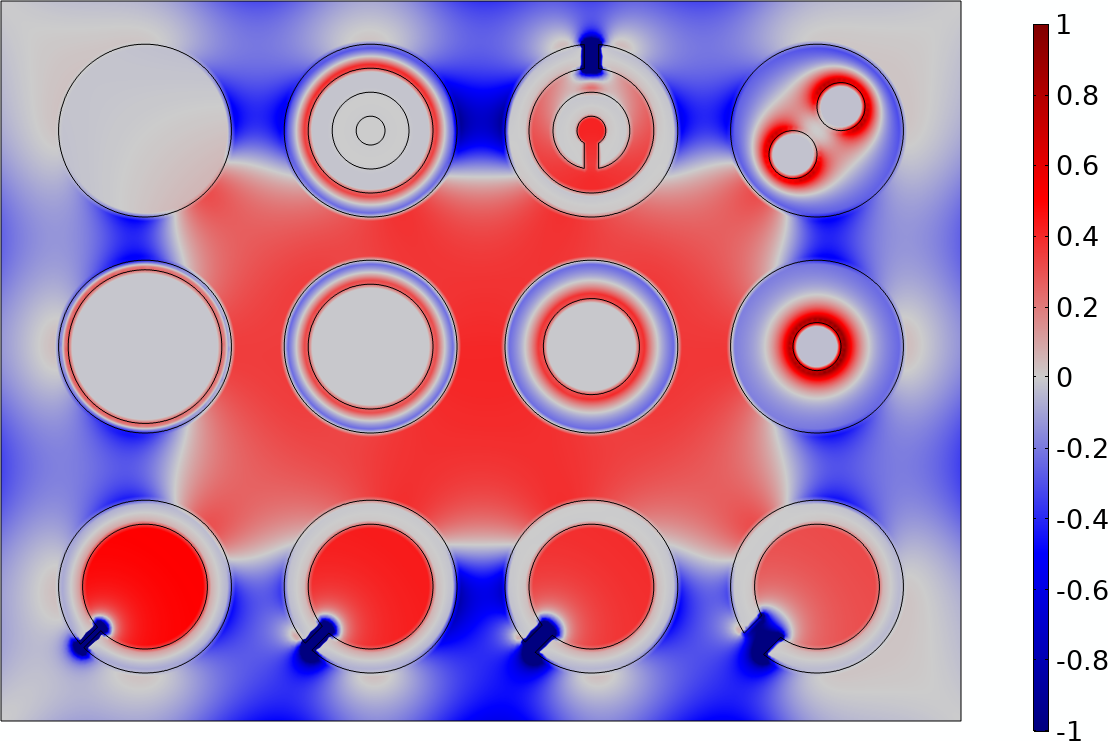}}
		\caption{Topological derivative of the constraint function $G_p$ for various values of the diffusion coefficient $a_p$}
		\label{fig:td-a}
	\end{center}
\end{figure*}
Figures \ref{fig:p-a} and \ref{fig:td-a} display the computed fictitious field $p$ and topological derivative with respect to the constraint function $G_p$.
As illustrated in Figure \ref{fig:p-a}, if the diffusion coefficient $a_p$ is set sufficiently large, the desired profile of $p$ is obtained. In other words, the value of $p$ is zero almost everywhere where the geometric constraint is satisfied, and it is not zero in the domains where the geometric constraint is violated.
This indicates that the definition of the constraint function $G_p$ is appropriate.
The distribution of the topological derivatives is similarly appropriate if the diffusion coefficient is sufficiently small, as illustrated in Figure \ref{fig:td-a}.
It should be noted that a small slit can be detected by adjusting the diffusion coefficient $a_p$. This signifies that the minimum size of a slit for powder removal can be qualitatively controlled by setting the diffusion coefficient. Additionally, since the design sensitivity contains information that is close to the constraint violation state, the effect is to suppress oscillation between the violation state and non-violation state in the process of the iterative calculation of the optimal design.

Next, we examine the appropriate range and effect of the parameter values with respect to the small parameter $\epsilon_p$, where the small diffusion coefficient $a_p$ is set to $1\times10^{2}$. 

\begin{figure*}[htb]
	\begin{center}
		\subfigure[$\epsilon_{p}=1\times10^{-6}$]{
			\includegraphics[height=2.5cm]{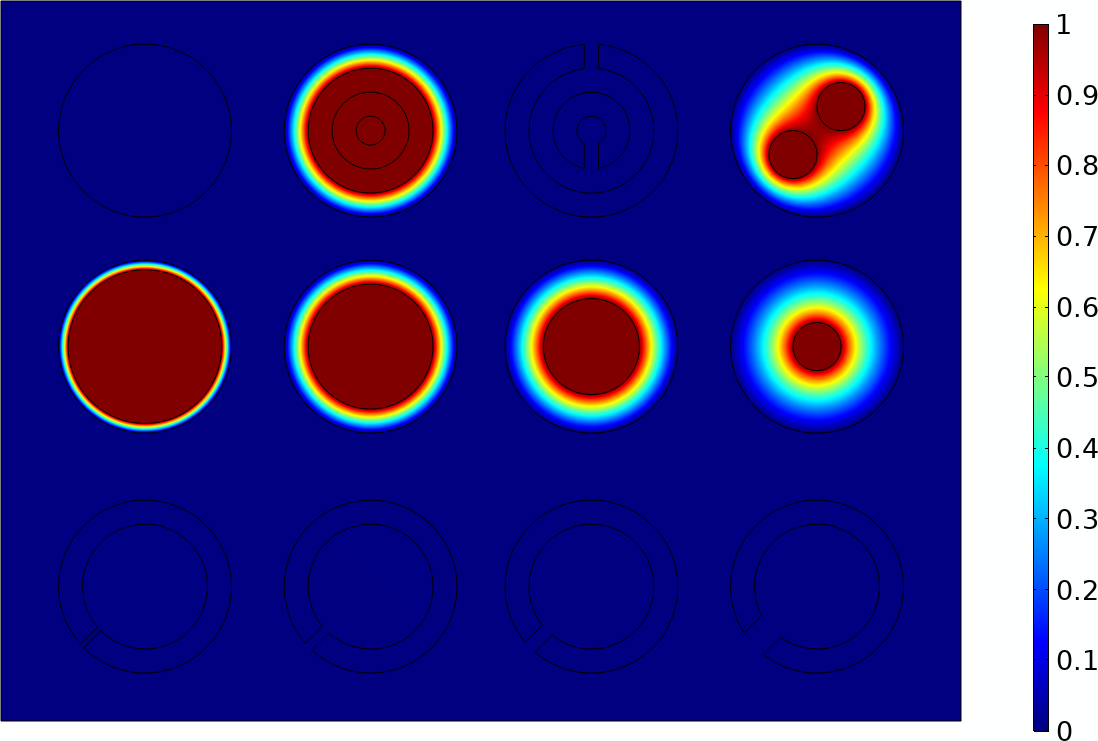}}
		\subfigure[$\epsilon_{p}=1\times10^{-5}$]{
			\includegraphics[height=2.5cm]{figure/p_1e2_1e-5_.png}}
		\subfigure[$\epsilon_{p}=1\times10^{-4}$]{
			\includegraphics[height=2.5cm]{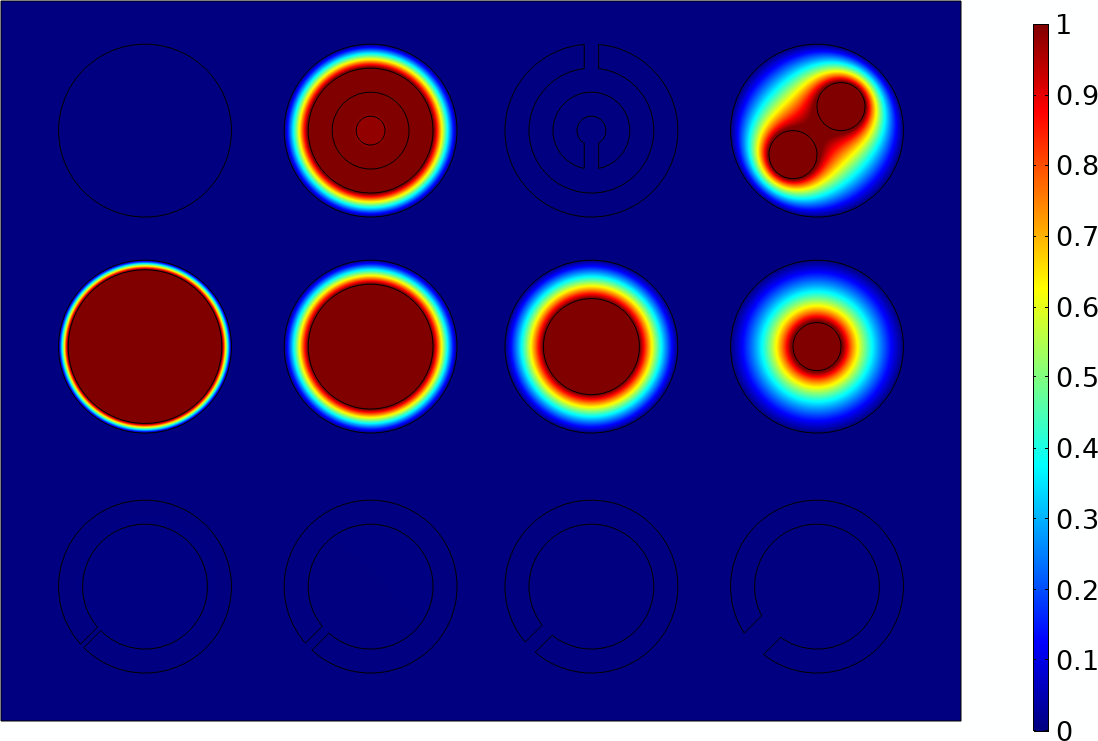}}
		\subfigure[$\epsilon_{p}=1\times10^{-3}$]{
			\includegraphics[height=2.5cm]{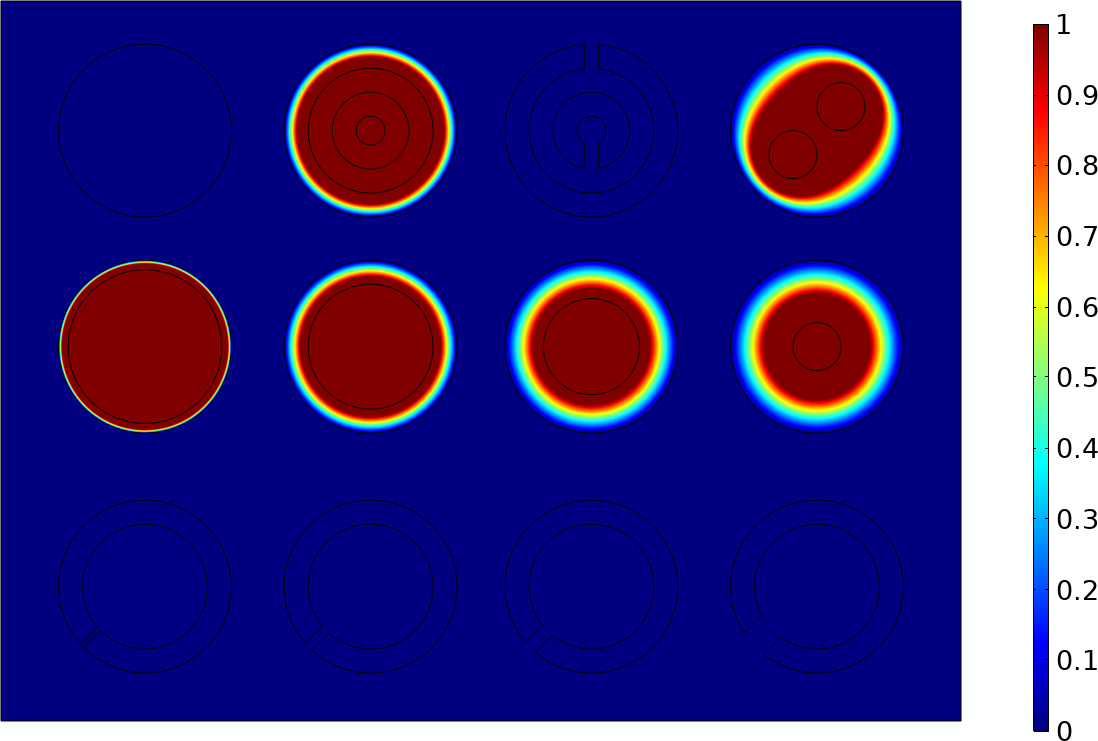}}
		\caption{The fictitious physical field $p$ for various setting of the relative small diffusion coefficient $\epsilon_p$}
		\label{fig:p-e}
	\end{center}
\end{figure*}
\begin{figure*}[htb]
	\begin{center}
		\subfigure[$\epsilon_{p}=1\times10^{-6}$]{
			\includegraphics[height=2.5cm]{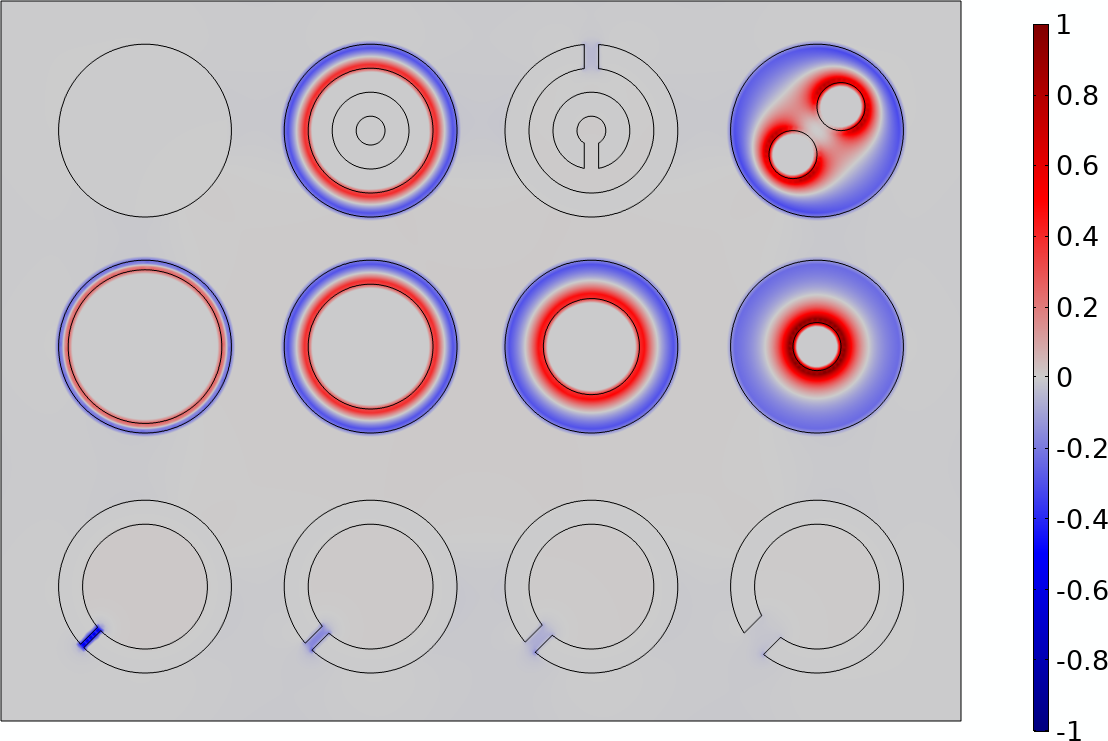}}
		\subfigure[$\epsilon_{p}=1\times10^{-5}$]{
			\includegraphics[height=2.5cm]{figure/td_1e2_1e-5_.png}}
		\subfigure[$\epsilon_{p}=1\times10^{-4}$]{
			\includegraphics[height=2.5cm]{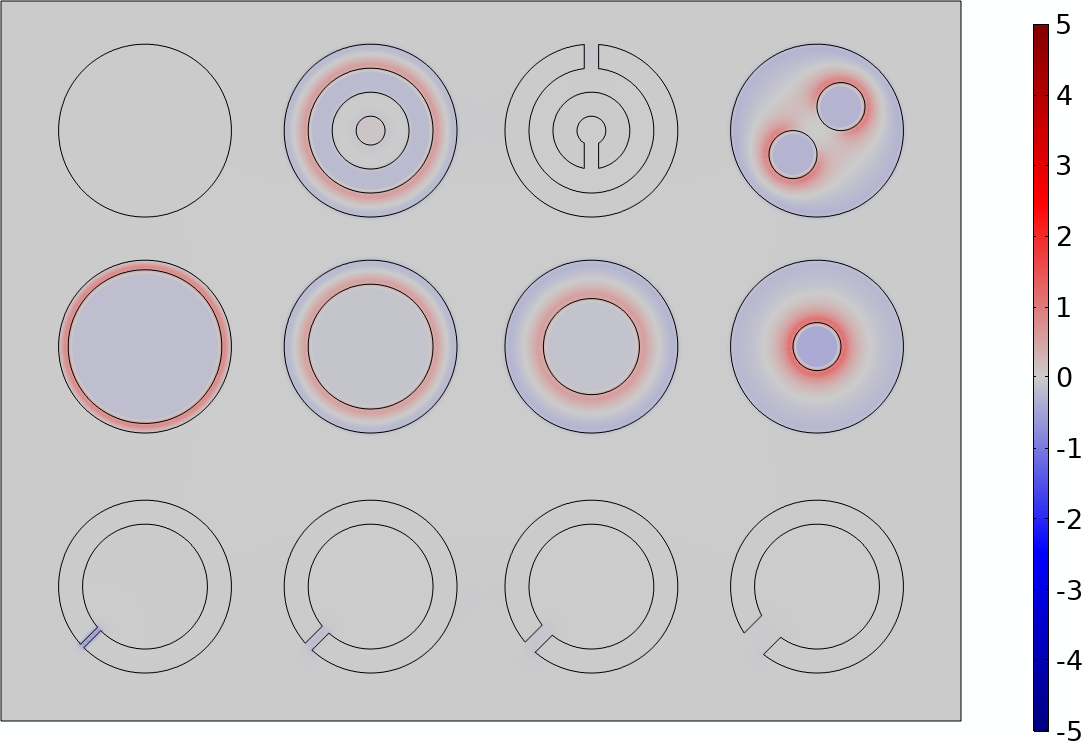}}
		\subfigure[$\epsilon_{p}=1\times10^{-3}$]{
			\includegraphics[height=2.5cm]{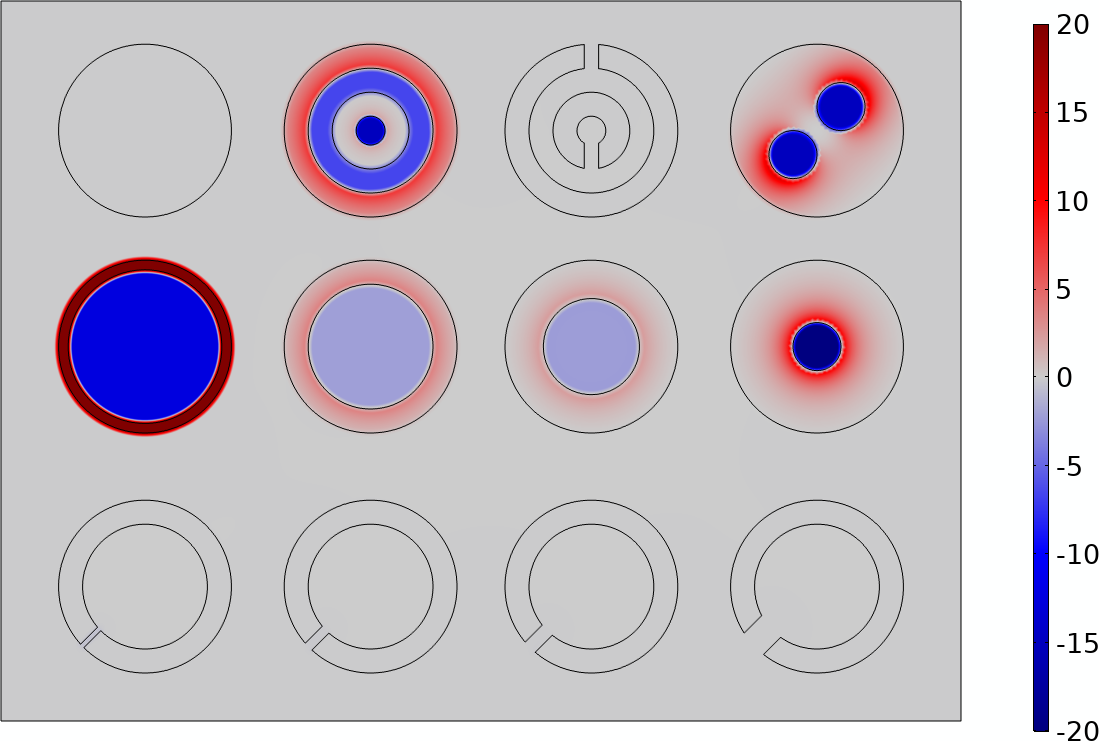}}
		\caption{The topological derivative of the constraint function $G_p$ for various setting of the relative small diffusion coefficient $\epsilon_p$}
		\label{fig:td-e}
	\end{center}
\end{figure*}
Figures \ref{fig:p-e} and \ref{fig:td-e} present the computed fictitious field $p$ and topological derivative with respect to the constraint function $G_p$.
As illustrated in Figure \ref{fig:td-e}, if the parameter $\epsilon_p$ is set to be smaller than $1\times 10^{-5}$, the topology derivatives are identical. Therefore, the parameter $\epsilon_p$ should be set to approximately $1\times 10^{-5}$. 

We focus on the four shapes placed at the bottom that satisfy the constraint owing to their small slits. The topological derivative is negative, suggesting that the slit should be slightly widened. To qualitatively control the width of the slit, the relationship between the size of the slit and the magnitude of the topological derivatives is important. Therefore, it is necessary to fix the parameter $\epsilon_p$ to $1\times 10^{-5}$ and study the relationship between parameter $a_p$ and the slit size in advance.

\begin{table}[htb]
	\caption{Parameter setting values in numerical examples}
	\label{tab:param}
	\centering
\begin{tabular}{lccc}
\hline
& $a_p$ & $\epsilon_p$ & result\\
\hline \hline
Fig. 3 (a) and Fig. 4 (a) \quad &$1\times 10^4$ & $1\times 10^{-5}$ & possible\\
Fig. 3 (b) and Fig. 4 (b) \quad &$1\times 10^3$ & $1\times 10^{-5}$ & appropriate\\
Fig. 3 (c) and Fig. 4 (c) \quad &$1\times 10^2$ & $1\times 10^{-5}$ & appropriate\\
Fig. 3 (d) and Fig. 4 (d) \quad &$1\times 10^1$ & $1\times 10^{-5}$ & inappropriate\\
Fig. 3 (e) and Fig. 4 (e) \quad &$1\times 10^0$ & $1\times 10^{-5}$ & inappropriate\\
\hline
Fig. 5 (a) and Fig. 6 (a) \quad &$1\times 10^2$ & $1\times 10^{-6}$ & appropriate\\
Fig. 5 (b) and Fig. 6 (b) \quad &$1\times 10^2$ & $1\times 10^{-5}$ & appropriate\\
Fig. 5 (c) and Fig. 6 (c) \quad &$1\times 10^2$ & $1\times 10^{-4}$ & inappropriate\\
Fig. 5 (d) and Fig. 6 (d) \quad &$1\times 10^2$ & $1\times 10^{-3}$ & inappropriate\\
\hline
\end{tabular}
\end{table}
Table \ref{tab:param} lists the above mentioned parameters. $\epsilon_p$ should be set to a value of $\le 1\times 10^{-5}$. $a_p$ can express the desired constraint if its value is $\ge 1\times 10^2$. This is important information for the shape-finding process to express the cases where the constraints are met, but a slight change in shape can result in a shape that does not meet the constraints as well as the cases where the constraints are not met, but a slight change in shape can result in a shape that meets the constraints. To retain this information, $a_p$ should be set to a value between $1\times 10^2$ and $1\times 10^3$. Even if the sizes of the design domain, etc. are changed, the range of these parameter settings will always be valid because the fictitious physical models are dimensionless.

\subsection{Minimum mean compliance problem}
\begin{figure*}[htb]
	\begin{center}
		\includegraphics[height=4cm]{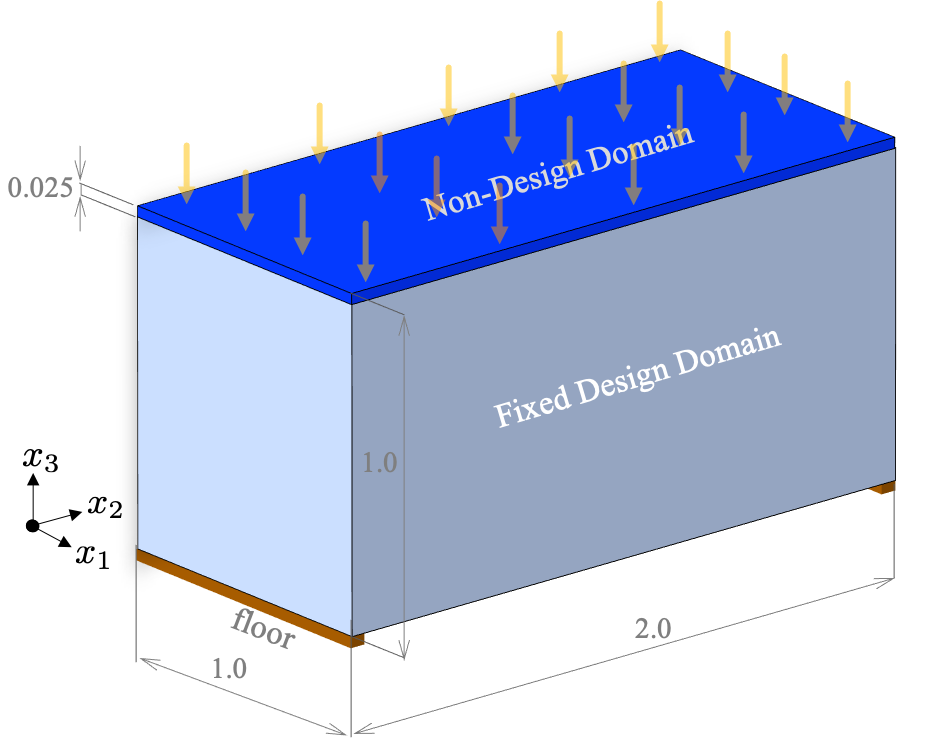}
		\caption{Problem setting for minimum mean compliance problem}
		\label{fig:fdd-table}
	\end{center}
\end{figure*}

\begin{figure*}[htb]
	\begin{center}
		\subfigure[Configuration; $\tau=5\times10^{-5}$]{
			\includegraphics[width=3cm]{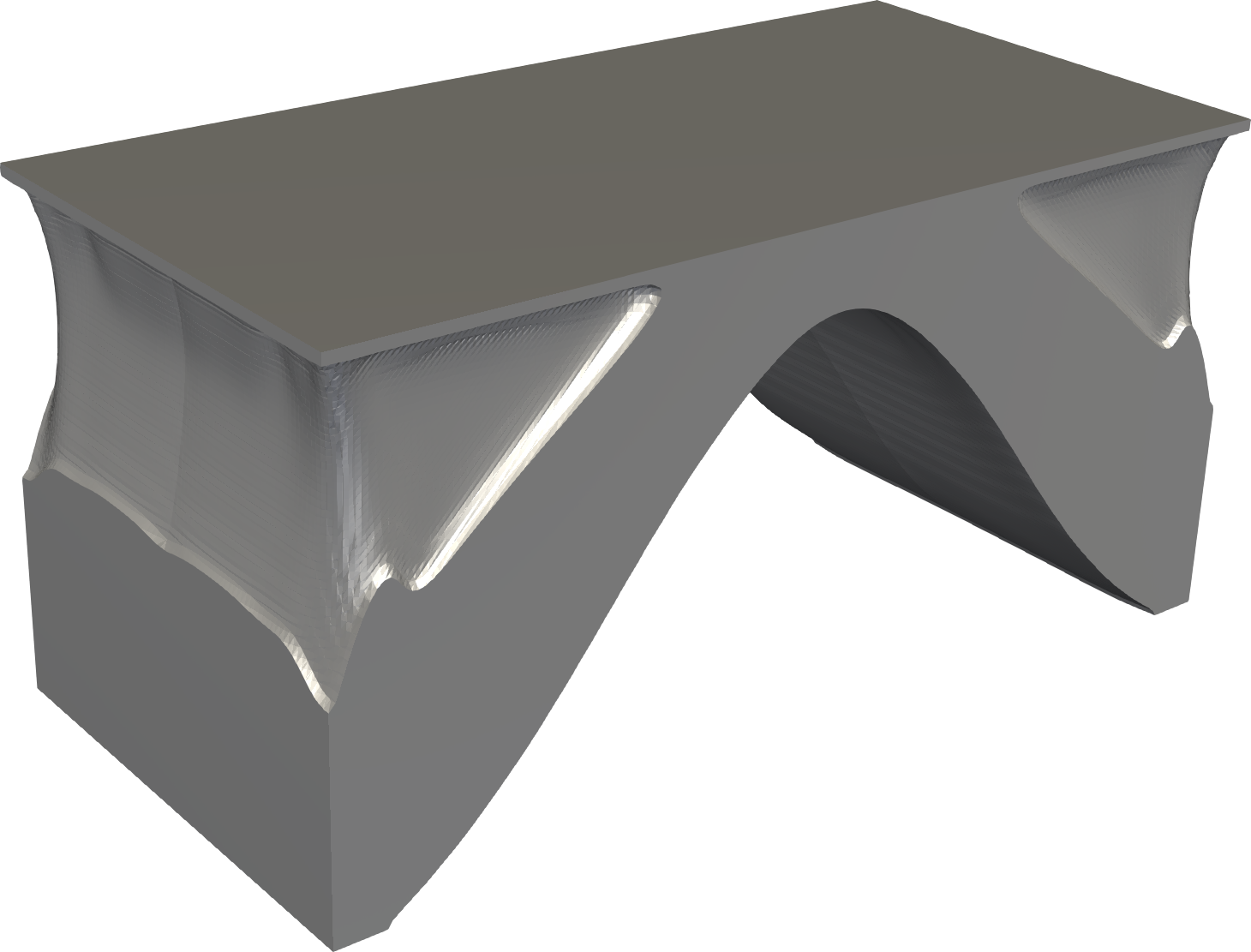}
		}\,
		\subfigure[Configuration; $\tau=1\times10^{-4}$]{
			\includegraphics[width=3cm]{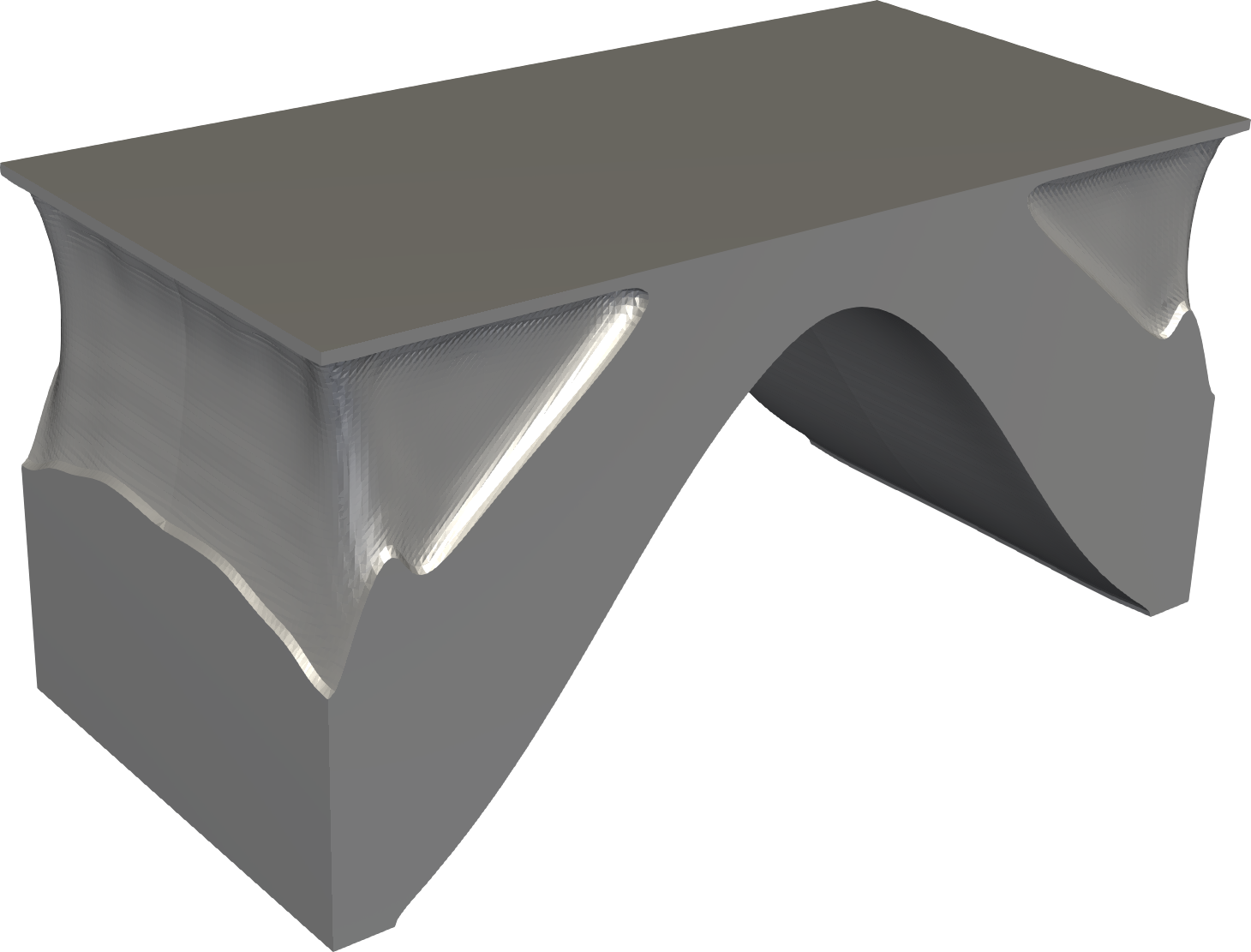}
		}\,
		\subfigure[Configuration; $\tau=5\times10^{-4}$]{
			\includegraphics[width=3cm]{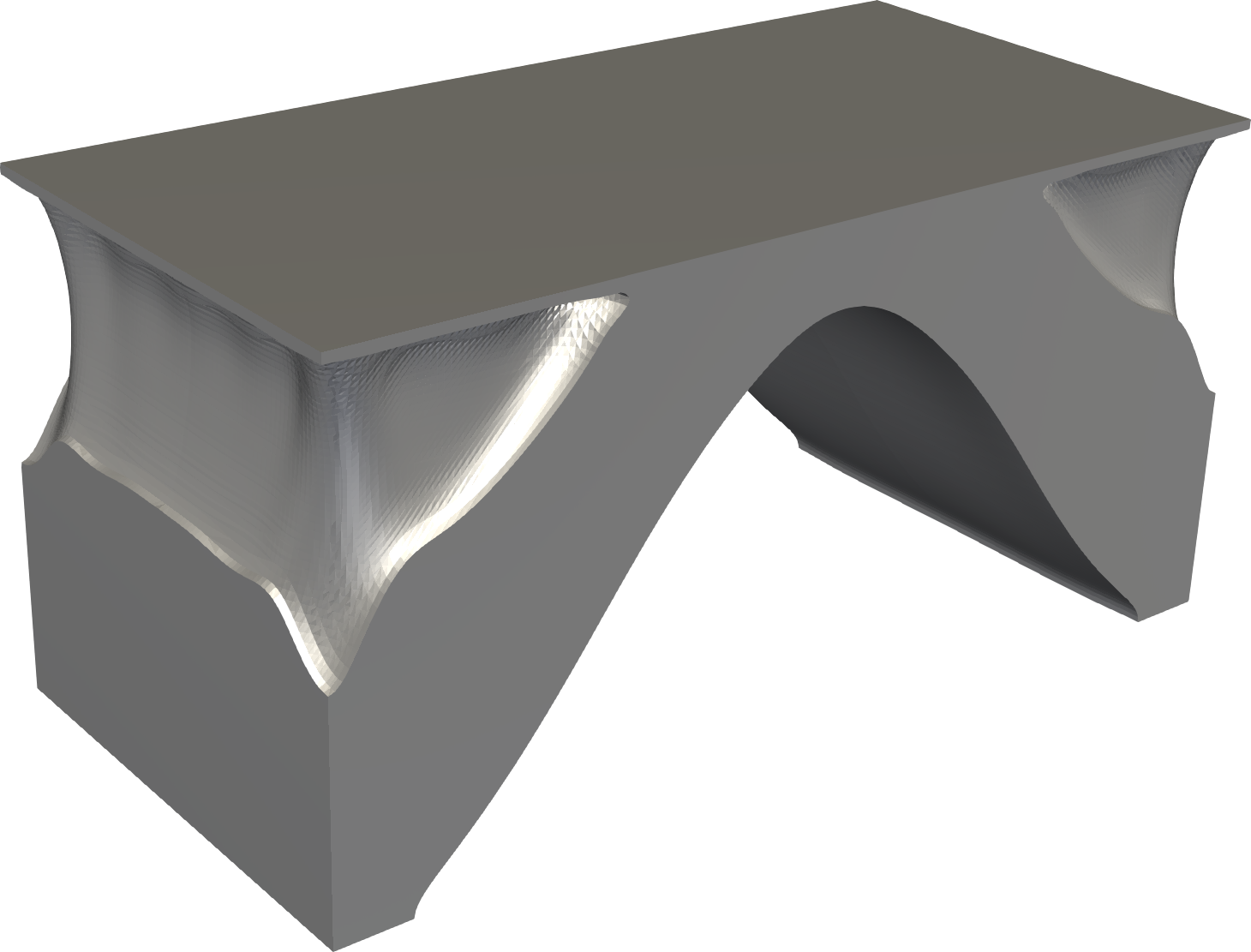}
		}\\
		\subfigure[Cross-sectional configuration; $\tau=5\times10^{-5}$]{
			\includegraphics[width=3cm]{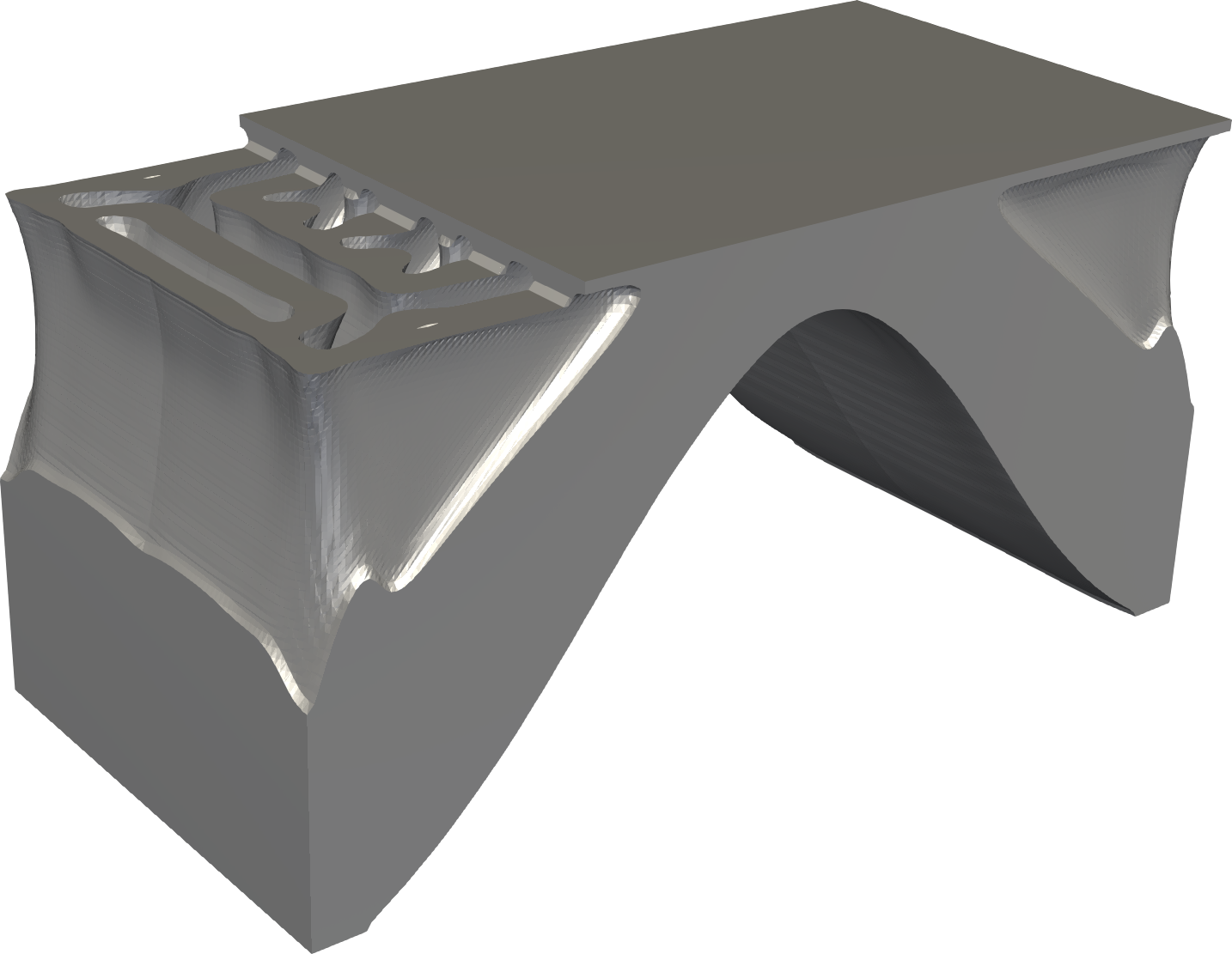}
		}\,
		\subfigure[Cross-sectional configuration; $\tau=1\times10^{-4}$]{
			\includegraphics[width=3cm]{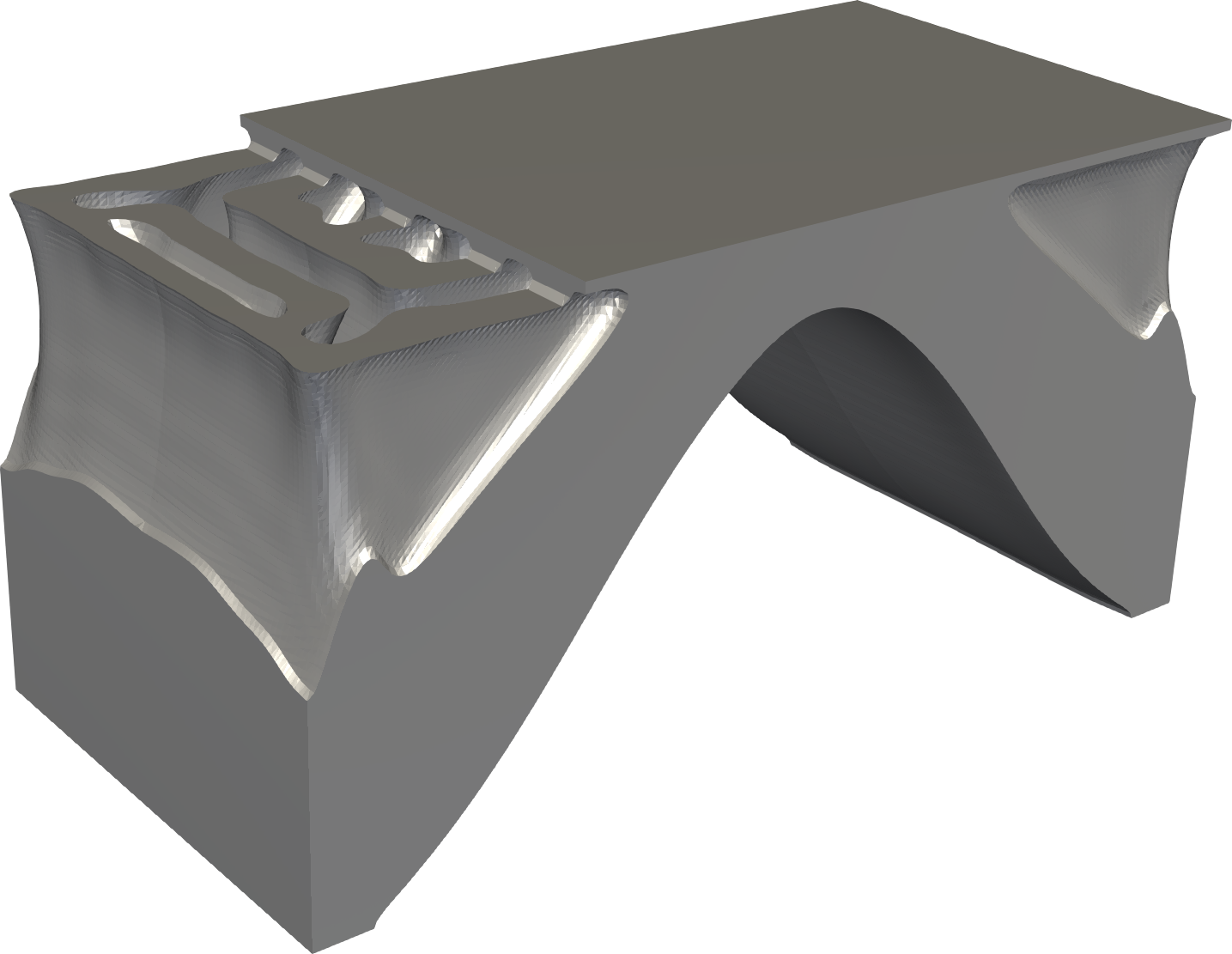}
		}\,
		\subfigure[Cross-sectional configuration; $\tau=5\times10^{-4}$]{
			\includegraphics[width=3cm]{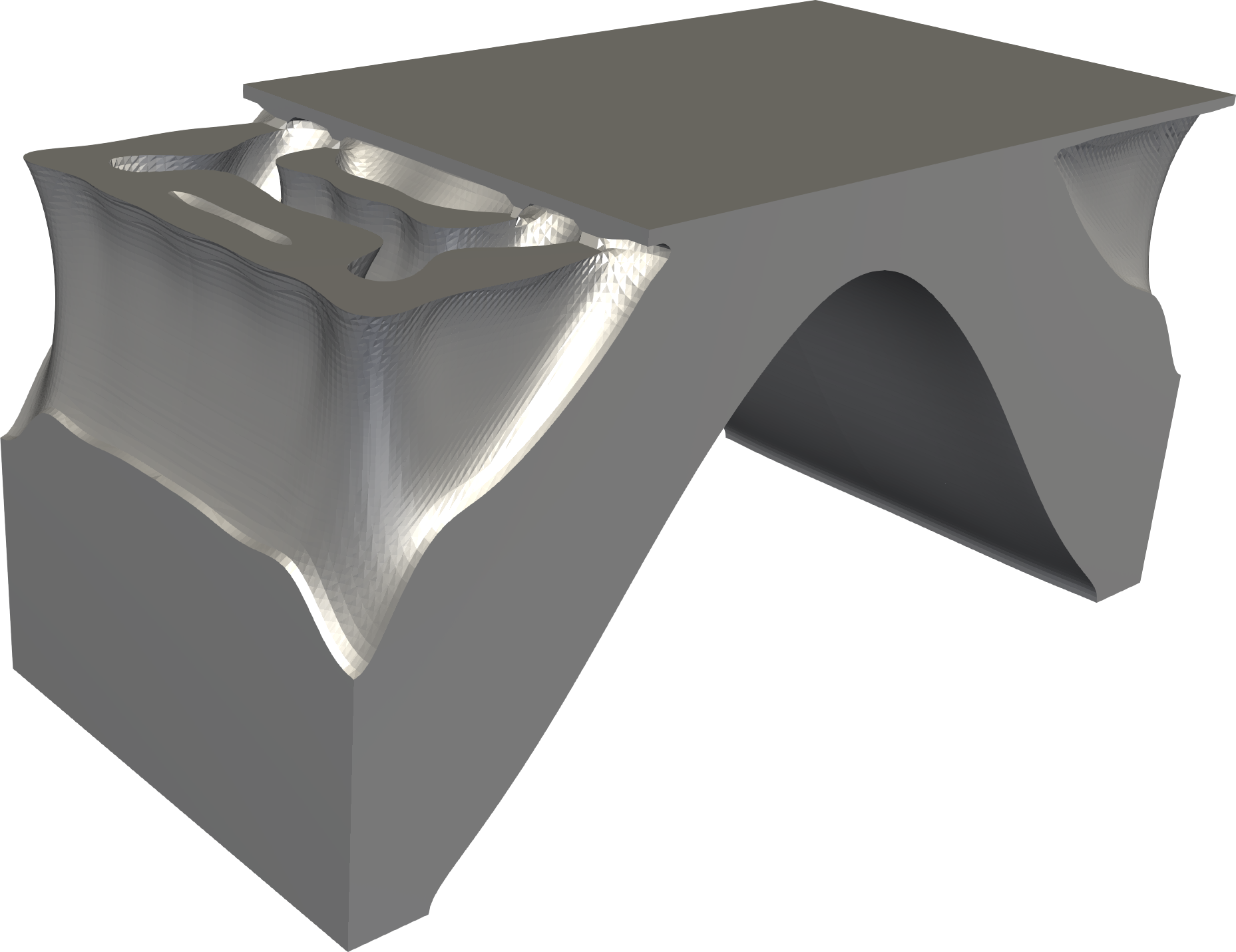}
		}
		\caption{Case 1: optimal configuration and its cross-sectional configuration without geometric constraint}
		\label{fig:tab-without}
	\end{center}
\end{figure*}

\begin{figure*}[htb]
	\begin{center}
		\subfigure[Configuration; $\tau=5\times10^{-5}$]{
			\includegraphics[width=3cm]{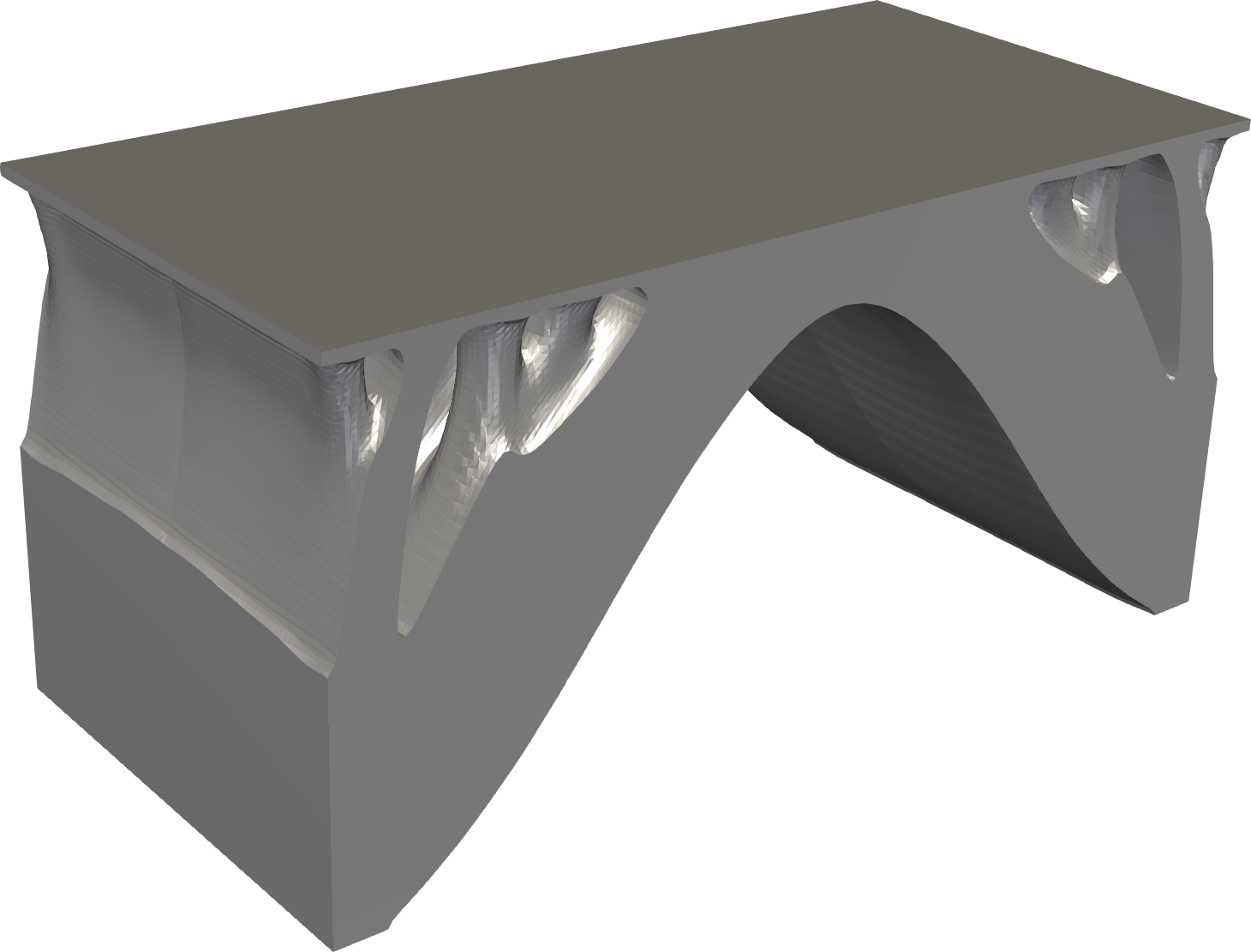}
		}\,
		\subfigure[Configuration; $\tau=1\times10^{-4}$]{
			\includegraphics[width=3cm]{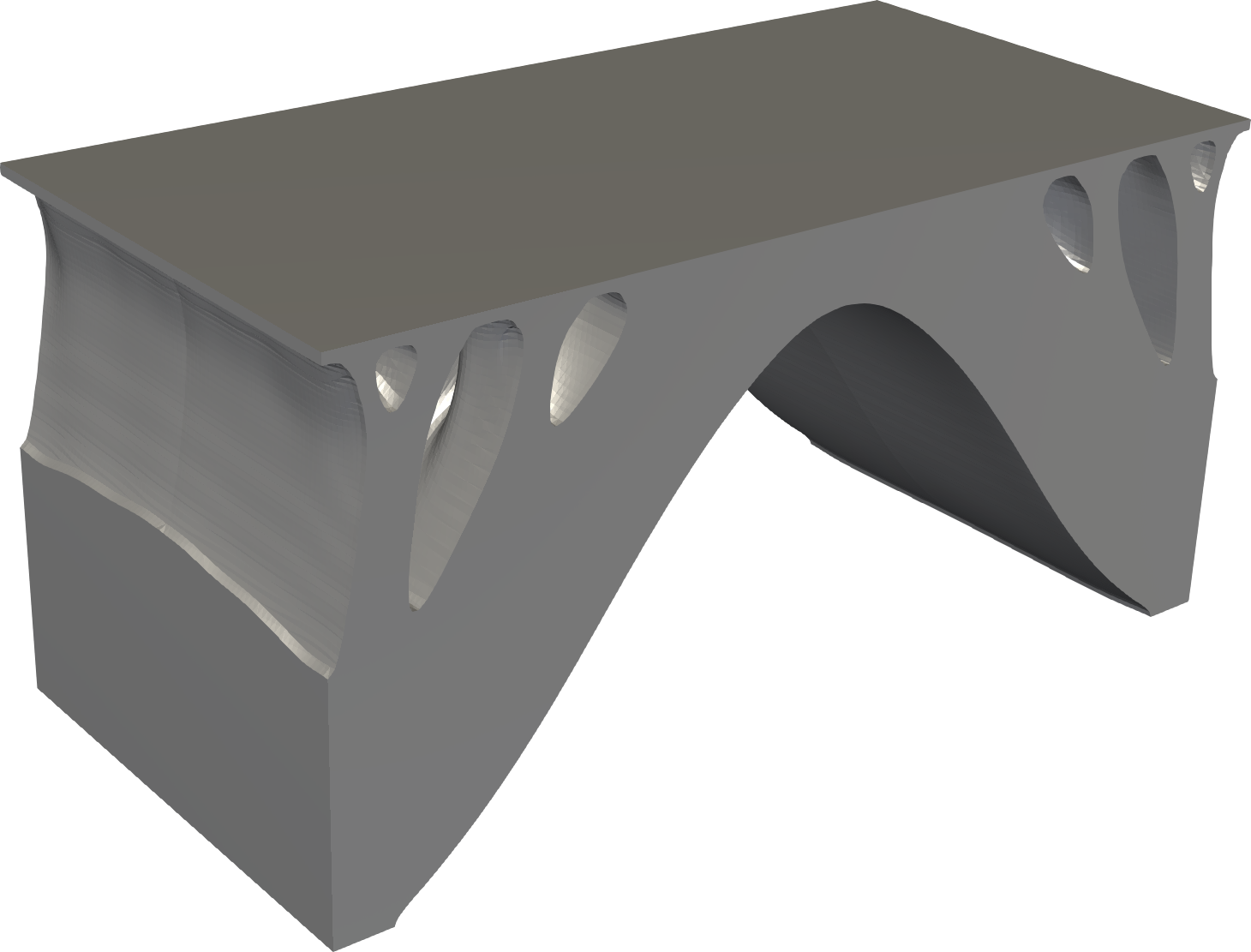}
		}\,
		\subfigure[Configuration; $\tau=5\times10^{-4}$]{
			\includegraphics[width=3cm]{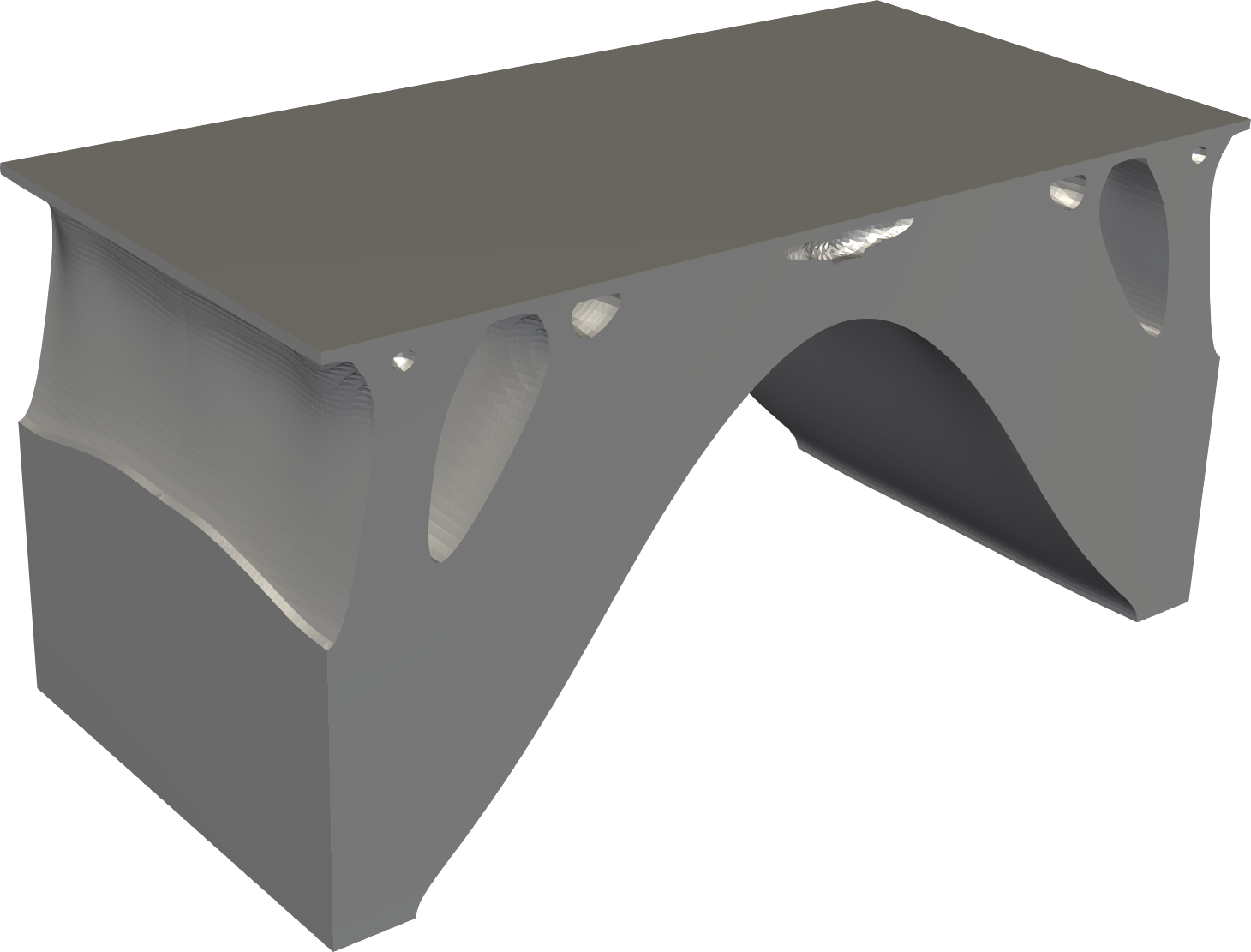}
		}\\
		\subfigure[Cross-sectional configuration; $\tau=5\times10^{-5}$]{
			\includegraphics[width=3cm]{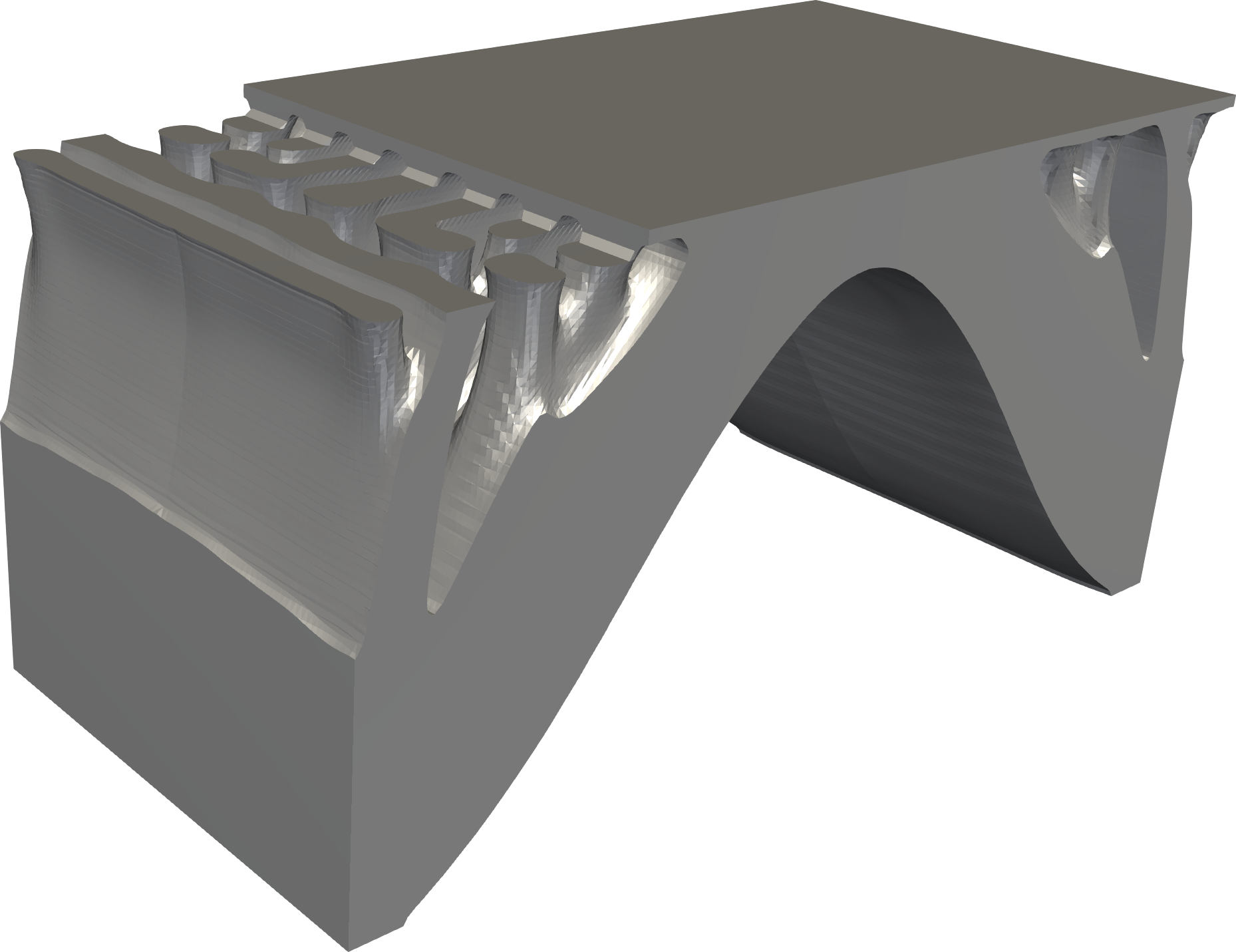}
		}\,
		\subfigure[Cross-sectional configuration; $\tau=1\times10^{-4}$]{
			\includegraphics[width=3cm]{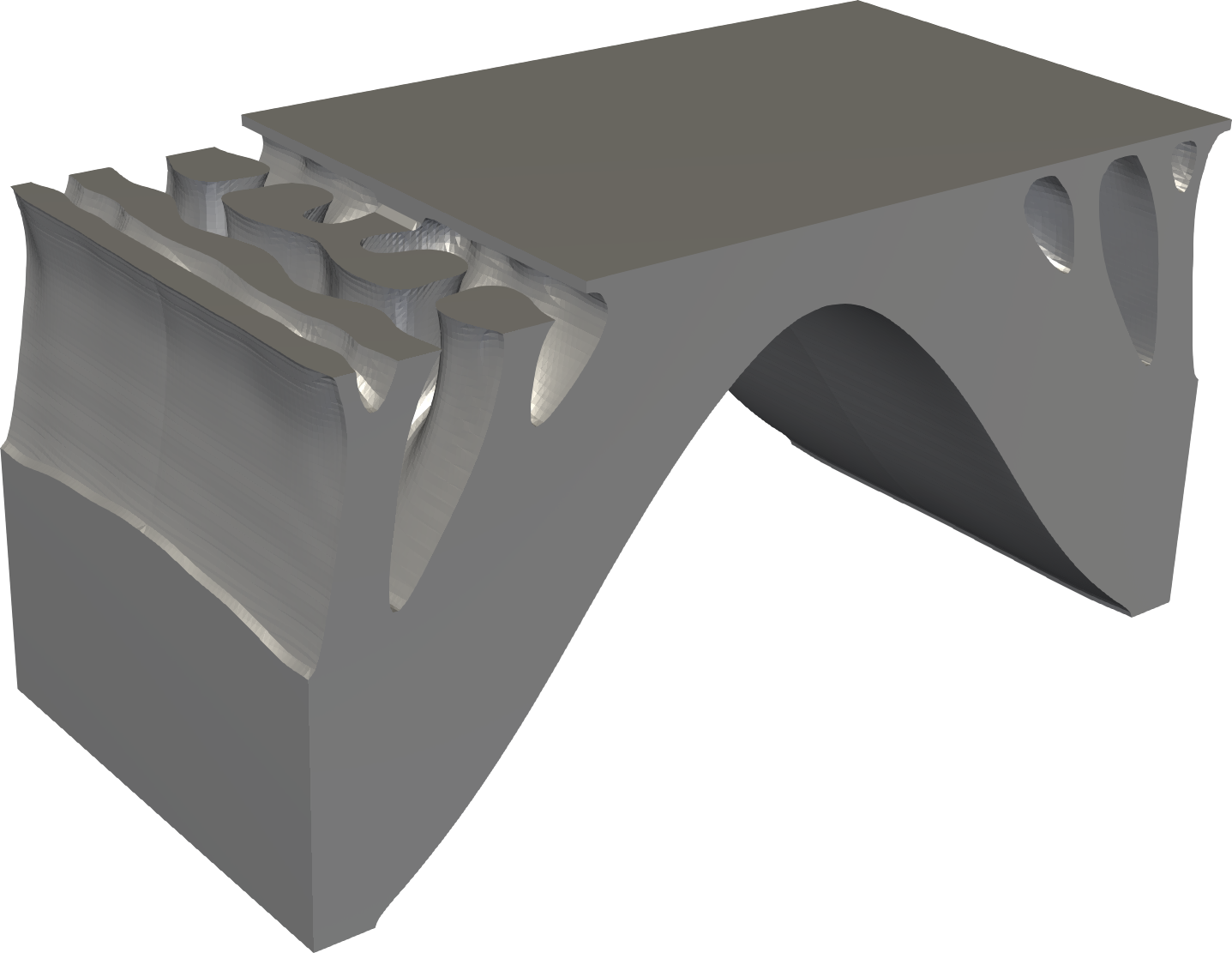}
		}\,
		\subfigure[Cross-sectional configuration; $\tau=5\times10^{-4}$]{
			\includegraphics[width=3cm]{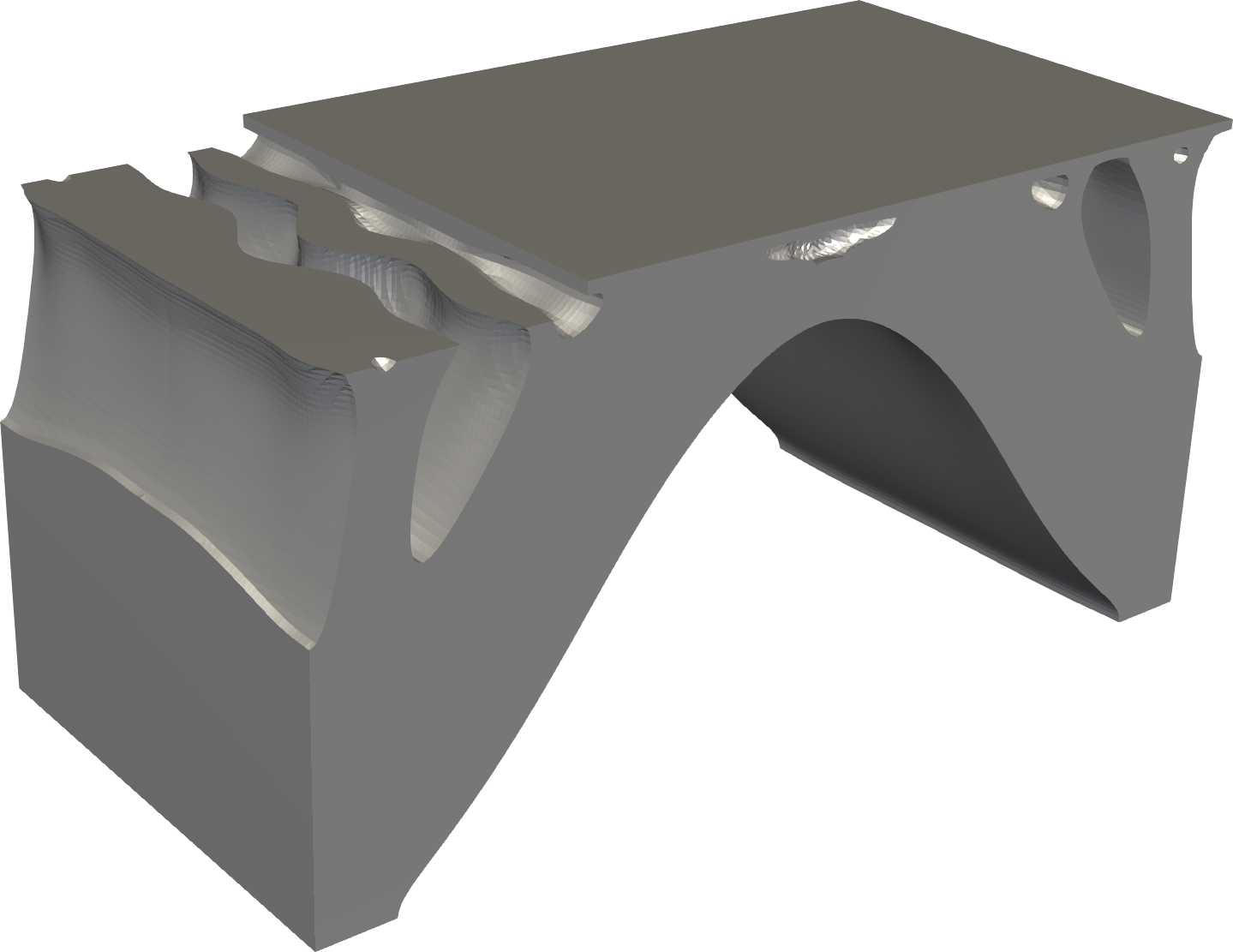}
		}
		\caption{Case 2: optimal configuration and its cross-sectional configuration considering the geometric constraint}
		\label{fig:tab-with}
	\end{center}
\end{figure*}

Here, optimization examples of the minimum mean compliance problem are presented to demonstrate the utility of the proposed method. 
In this section, the international system of units is used. The size of fixed design domain is set to a size of $1.0\rm{m}\times 2.0\rm{m} \times 1.0\rm{m}$.
Figure \ref{fig:fdd-table} presents the fixed design domain and boundary conditions. \\
\\
As can be seen, the non-design domain is introduced on the fixed design domain, and traction is applied to the top surface of the non-design domain. The fixed design domain is fixed at the left and right sides of the bottom surface. In addition, the domain is discretized into tetrahedral elements.
In these examples, the isotropic linear elastic material has Young's modulus of $210\, {\rm GPa}$ and Poisson ratio of $0.3$.
The diffusion coefficients relative to fictitious physical field $p$ are set to $a_p=1\times10^{1}$ and $\epsilon_{p}=1\times10^{-4}$.
Here, we examine the effect that the geometric constraint has on the resulting configurations. Figures \ref{fig:tab-without} and \ref{fig:tab-with} present the obtained optimal configurations without and with the geometric constraint, (i.e., cases 1 and 2), respectively.

Although both cases demonstrate clear and smooth configurations, the configuration of case 1 includes a cavity that is not connected to the outside. Conversely, all void domains in the configuration of case 2 are connected to the outside; therefore, the configuration of case 2 does not require an additional manufacturing process to reduce the metal powder from the LPBF process.

Now, we compare the characteristics of the shapes. When geometric constraints are not considered, stiffness is guaranteed by the plate-like structure. However, when geometric constraints are taken into account, the columnar structure ensures stiffness. Therefore, the proposed method makes it possible to select an appropriate structural type in consideration of the geometric constraints.
That is, it is difficult to obtain the optimal shape considering the geometric constraint based on a shape that was obtained without considering the geometric constraint. This is because the basic characteristics of the shape differ with and without constraints. Therefore, modifying the obtained shape while considering the constraints will not result in a proper modification. Herein, the proposed method searches for the optimal shape while considering the constraints; thus, it obtains a shape that is suitable for additive manufacturing.

\begin{figure}[htb]
	\begin{center}
		\includegraphics[height=4cm]{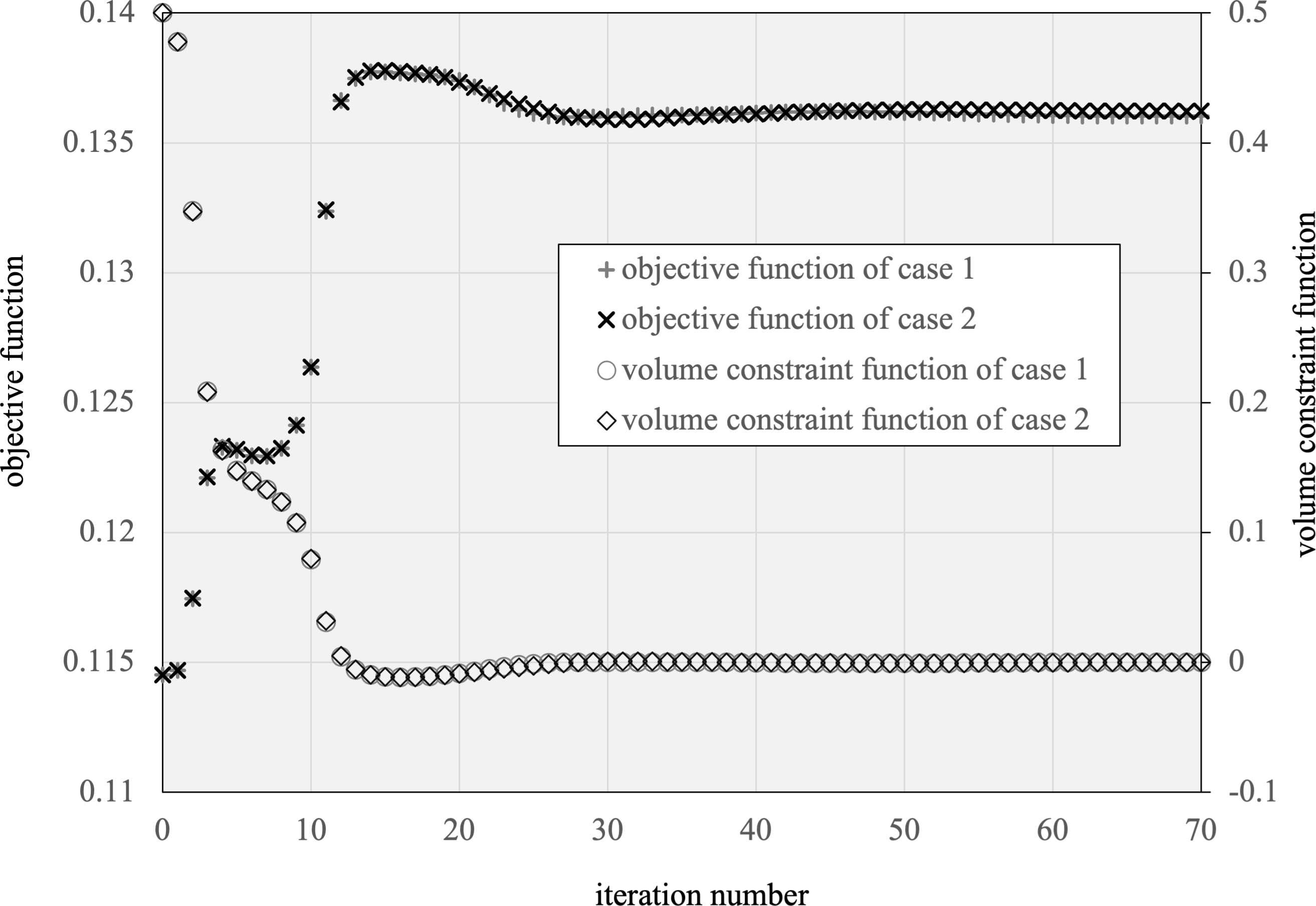}
		\caption{convergence histories of objective function and volume constraint function}
		\label{fig:plot}
	\end{center}
\end{figure}

Then, the convergence history and converged objective function value are compared.
Figure \ref{fig:plot} represents the convergence histories of objective function and volume constraint function in the case of $\tau=5\times10^{-5}$.
As shown in the figure, in both cases, the results converge after almost 50 iterations. 
The convergence is confirmed to not deteriorate when geometric constraints are added. The objective function value is slightly better when the geometric constraint is not considered. From the perspective of engineering applications, it is important to obtain a manufacturable shape even if the value of the objective function is slightly low.
Table \ref{tab:obj} presents the final objective function values, including the cases where the regularization parameter was set to $\tau=1\times10^{-4}$ and $\tau=5\times10^{-4}$.
\begin{table}[htb]
	\caption{Obtained objective function values}
	\label{tab:obj}
	\centering
\begin{tabular}{lcc}
\hline
regularization parameter \quad& objective function of case 1 \quad& objective function of case 2\\
\hline \hline
$\tau=5\times10^{-5}$ \quad &$0.13597$ & $0.13611$ \\
$\tau=1\times10^{-4}$ \quad &$0.13622$ & $0.13645$ \\
$\tau=5\times10^{-4}$ \quad &$0.13680$ & $0.13683$ \\
\hline
\end{tabular}
\end{table}
In all the cases, the difference in objective function values is minor. That is, the proposed method enables a design with high performance while satisfying the geometric constraints considering the manufacturing requirements.
\newpage
\subsection{Thermal diffusivity problem}
Next, optimization examples of the thermal diffusivity problem are presented to illustrate the utility of the proposed method. 
\\
\begin{figure*}[htb]
	\begin{center}
		\includegraphics[height=4cm]{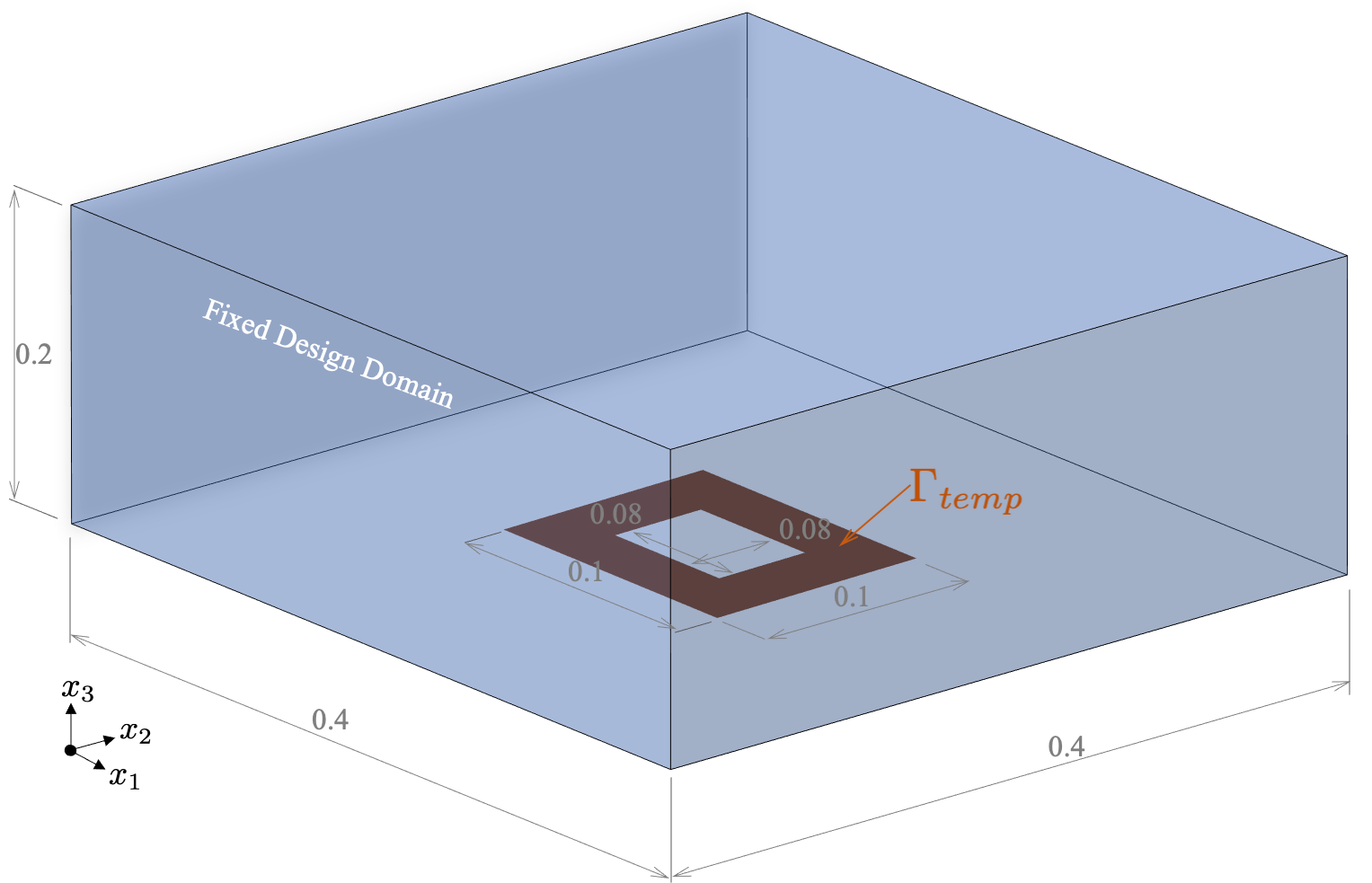}
		\caption{Problem setting for thermal diffusivity problem}
		\label{fig:fdd2}
	\end{center}
\end{figure*}\\
Figure \ref{fig:fdd2} presents the fixed design domain and boundary conditions.
As can be seen, the fixed design domain is rectangular with dimensions of $0.4 \text{ m}\times 0.4 \text{ m} \times 0.2 \text{ m}$ with heat source $Q$. The boundary $\Gamma_{temp}$ is prescribed to a relative temperature of $0$ K.
All other boundaries are subject to adiabatic boundary conditions.
The domain is discretized into tetrahedral elements.
In these examples, the thermal conductivities of the material and air are set to $26 \text{ W/mK}$ and $2.23\times 10^{-2} \text{ W/mK}$.
The diffusion coefficients relative to fictitious physical field $p$ are set to $a_p=1\times10^{2}$ and $\epsilon_{p}=1\times10^{-6}$, and the characteristic length is set to $L = 0.2$. The boundary $\Gamma_p$ is set to the side and top boundaries.
 
\begin{figure*}[htb]
	\begin{center}
		\subfigure[bird's-eye view; $\tau=1\times10^{-5}$]{
			\includegraphics[width=5cm]{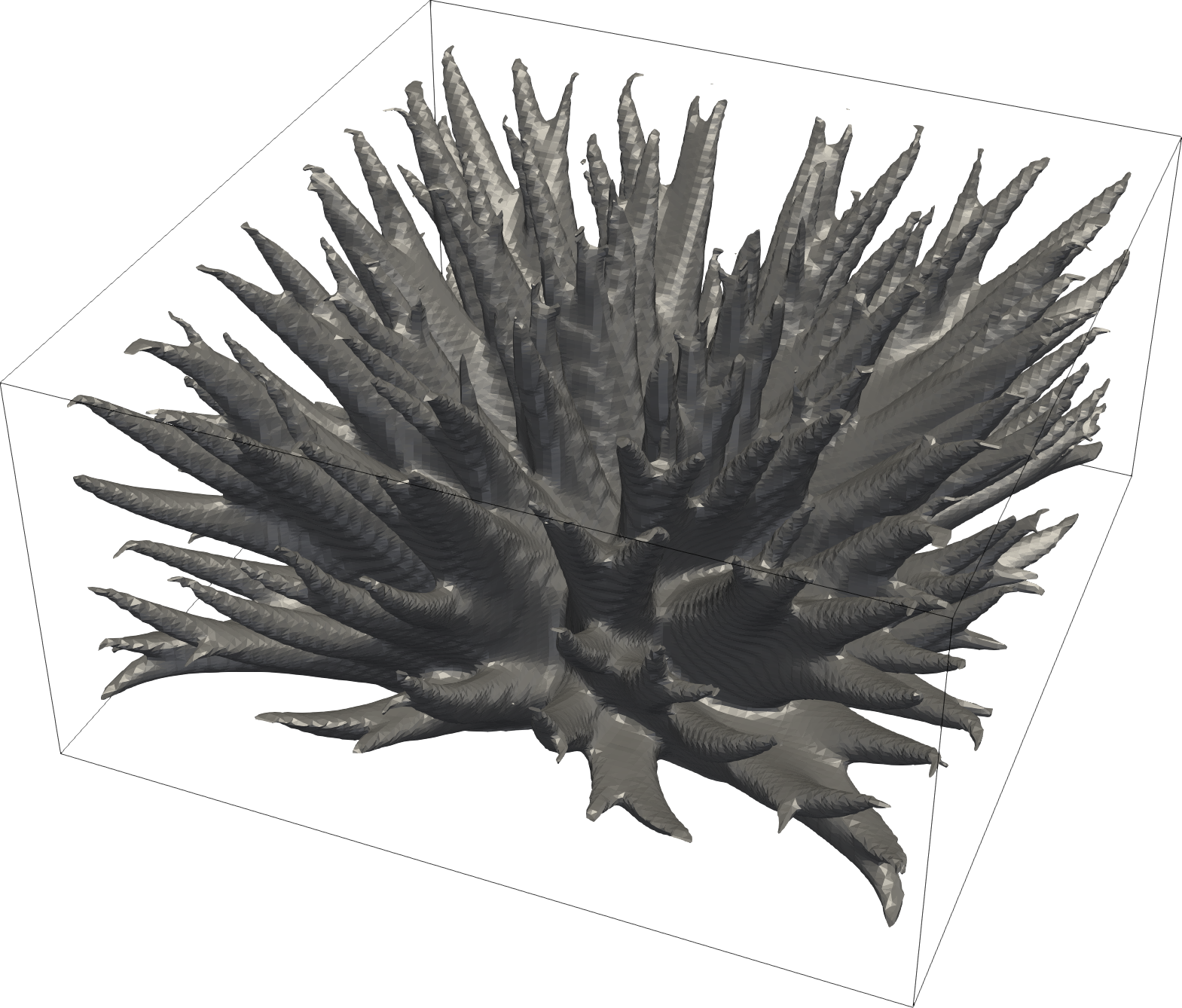}
		}\,
		\subfigure[bird's-eye view; $\tau=5\times10^{-4}$]{
			\includegraphics[width=5cm]{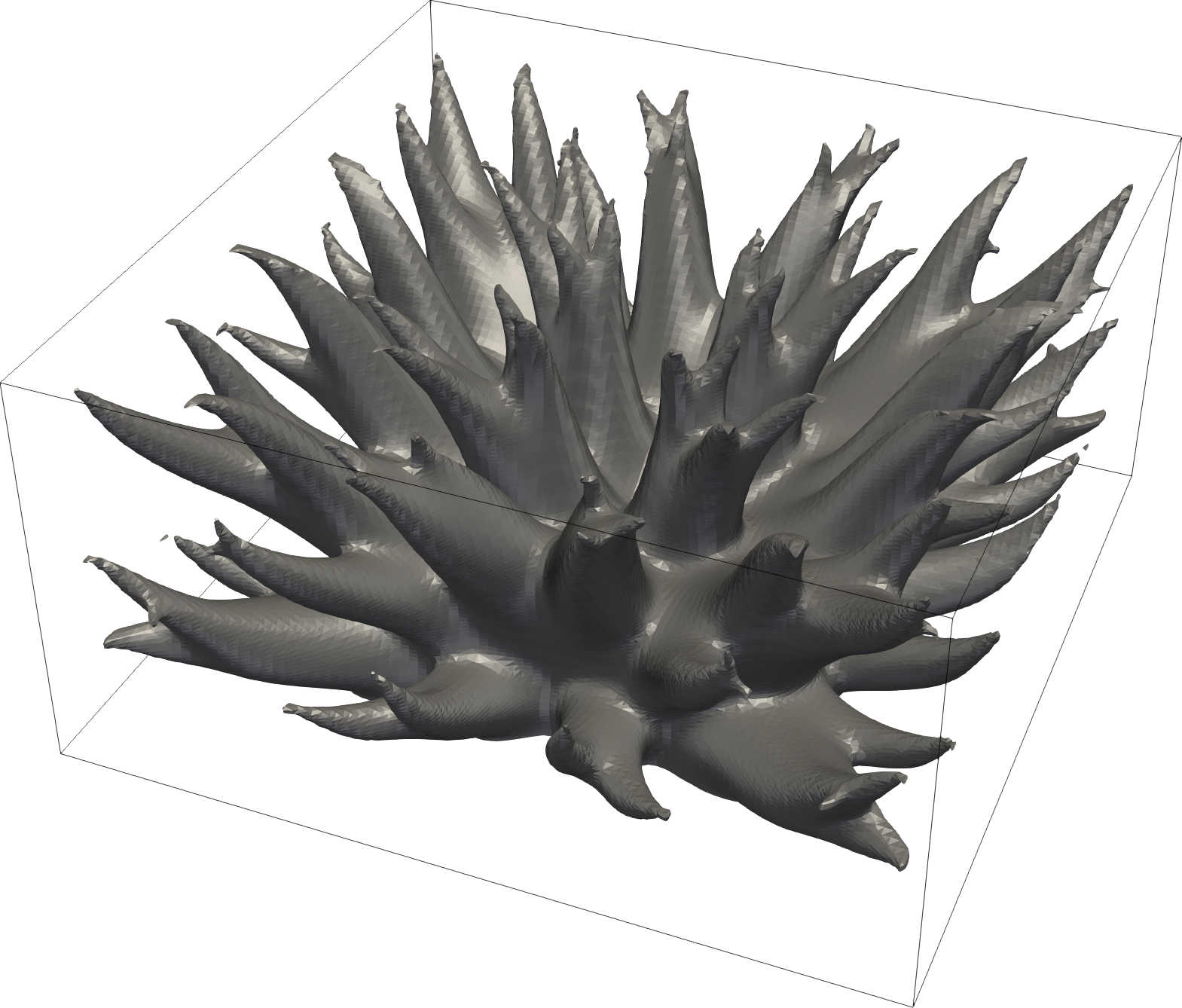}
		}\\
		\subfigure[bottom view; $\tau=1\times10^{-5}$]{
			\includegraphics[width=5cm]{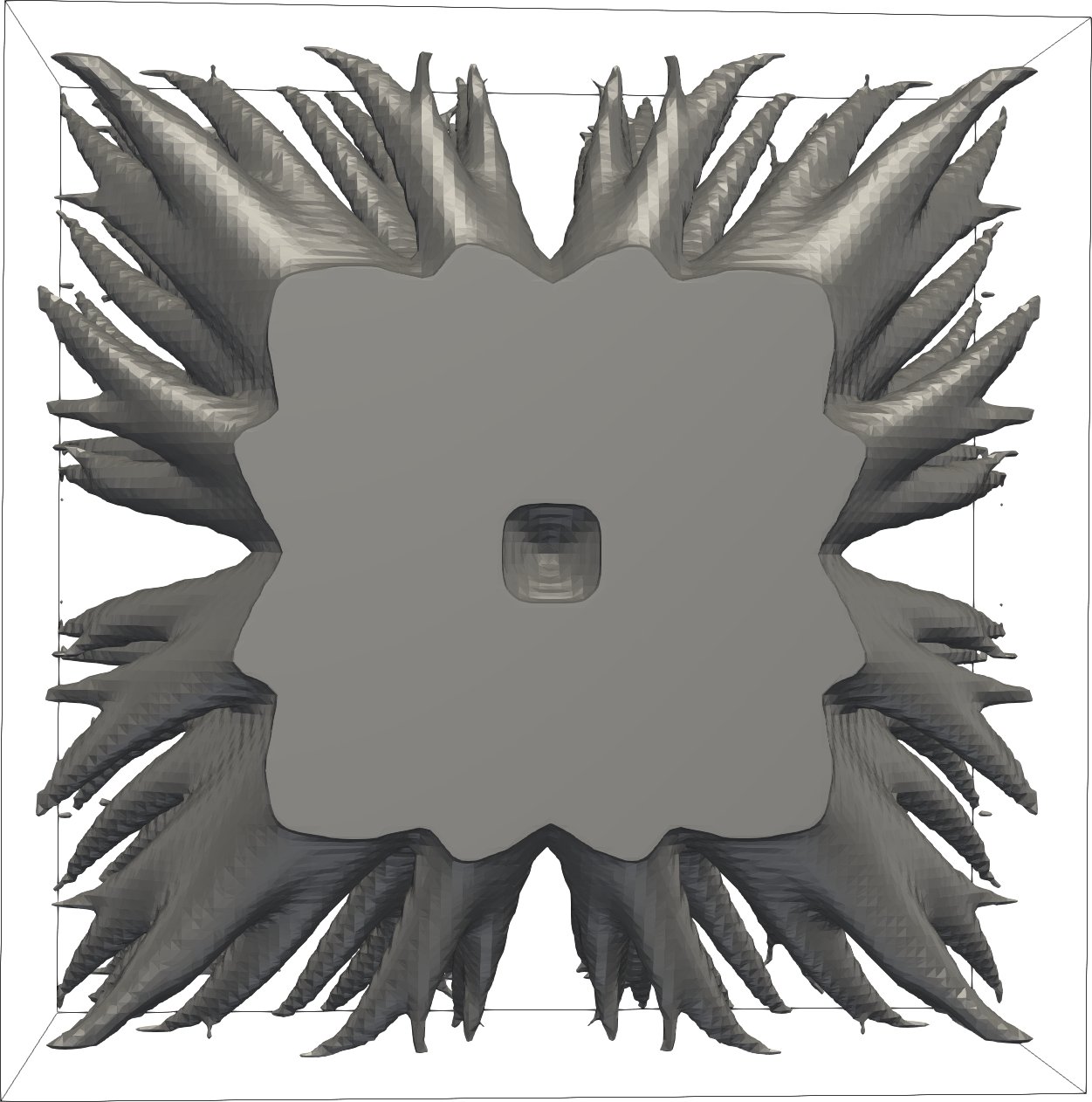}
		}\qquad \qquad
		\subfigure[bottom view; $\tau=5\times10^{-4}$]{
			\includegraphics[width=5cm]{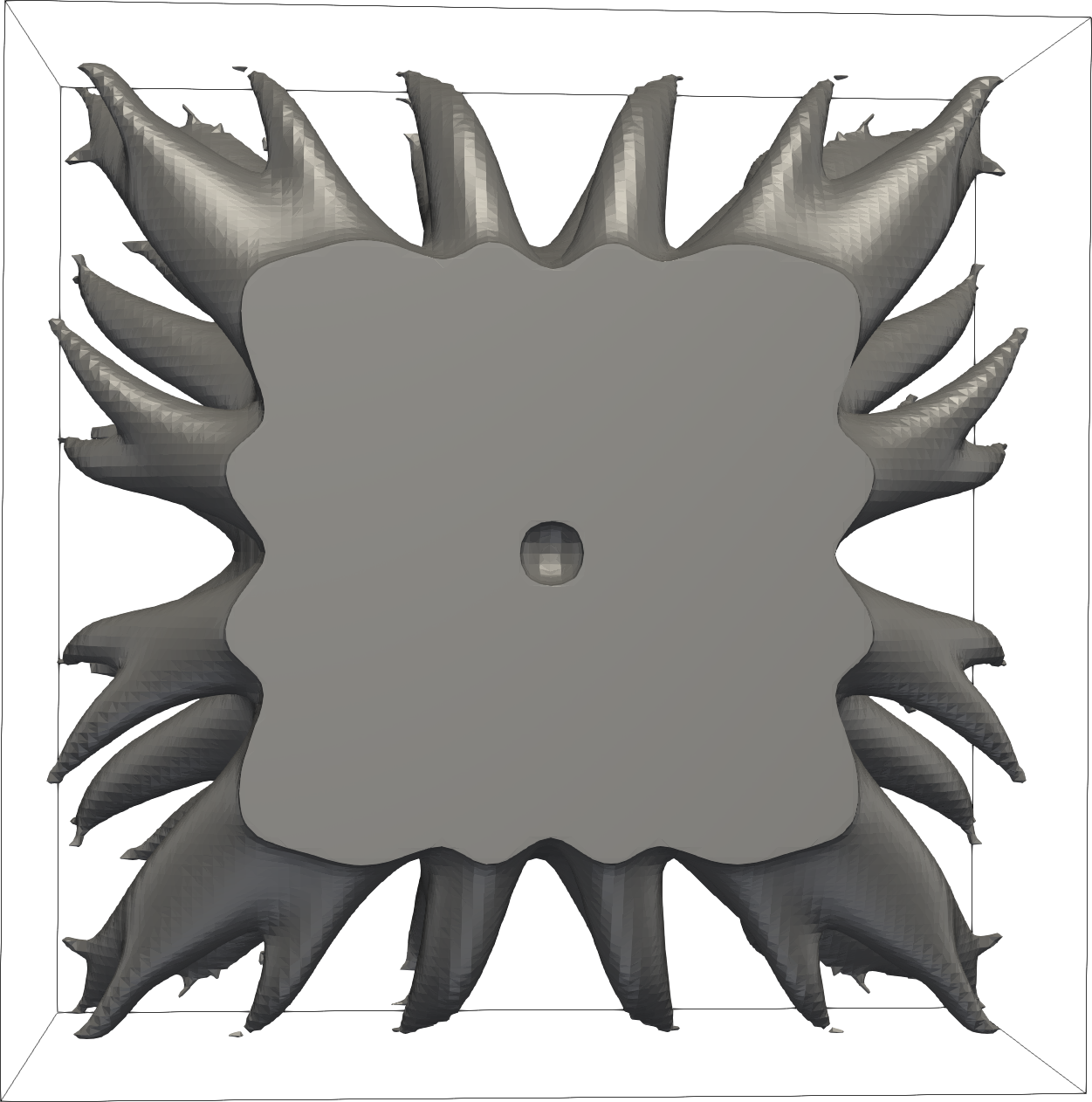}
		}
		\caption{Optimal configuration without geometrical constraint}
		\label{fig:thermal-without}
	\end{center}
\end{figure*}

\begin{figure*}[htb]
	\begin{center}
		\subfigure[bird's-eye view; $\tau=1\times10^{-5}$]{
			\includegraphics[width=5cm]{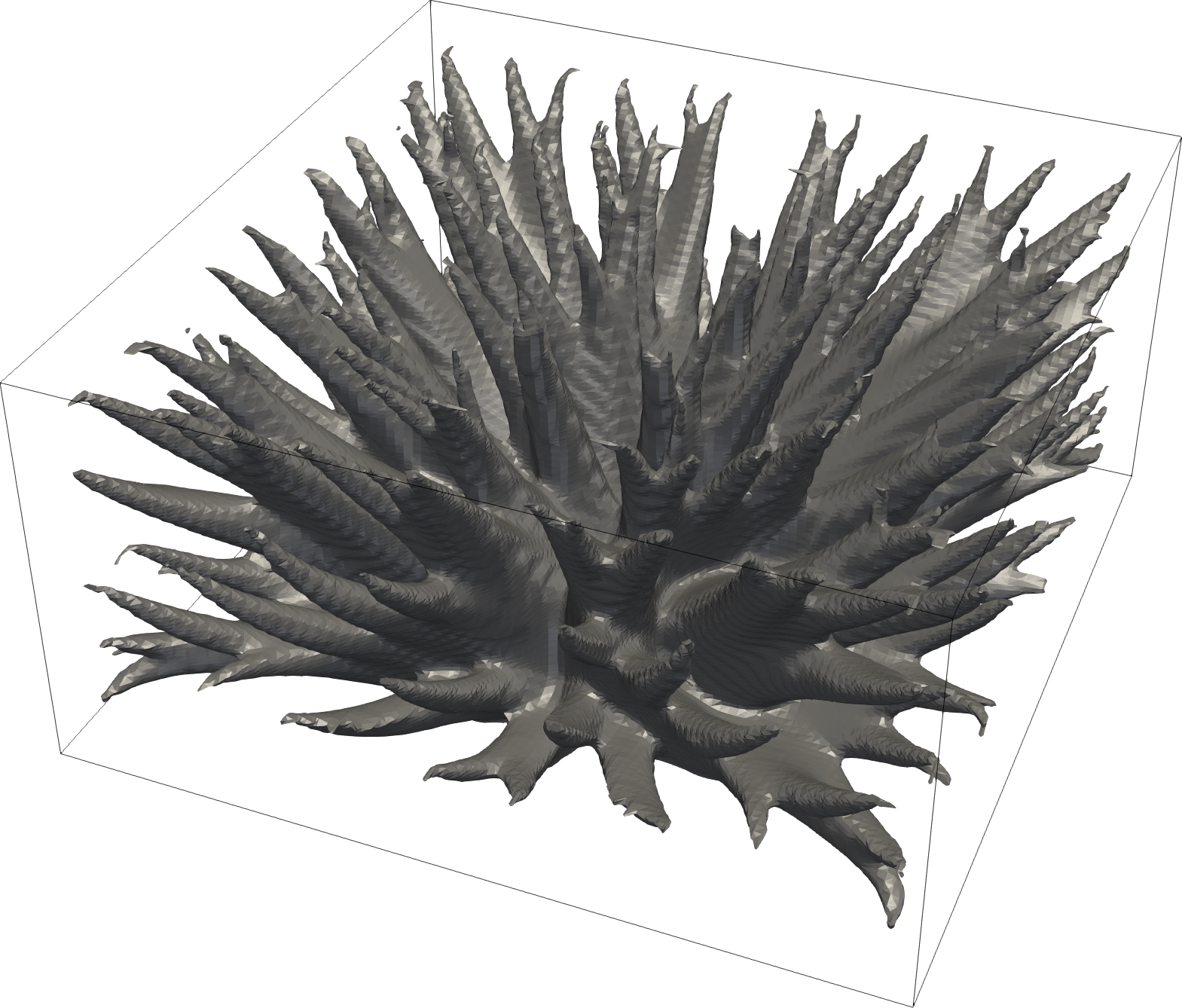}
		}\,
		\subfigure[bird's-eye view; $\tau=5\times10^{-4}$]{
			\includegraphics[width=5cm]{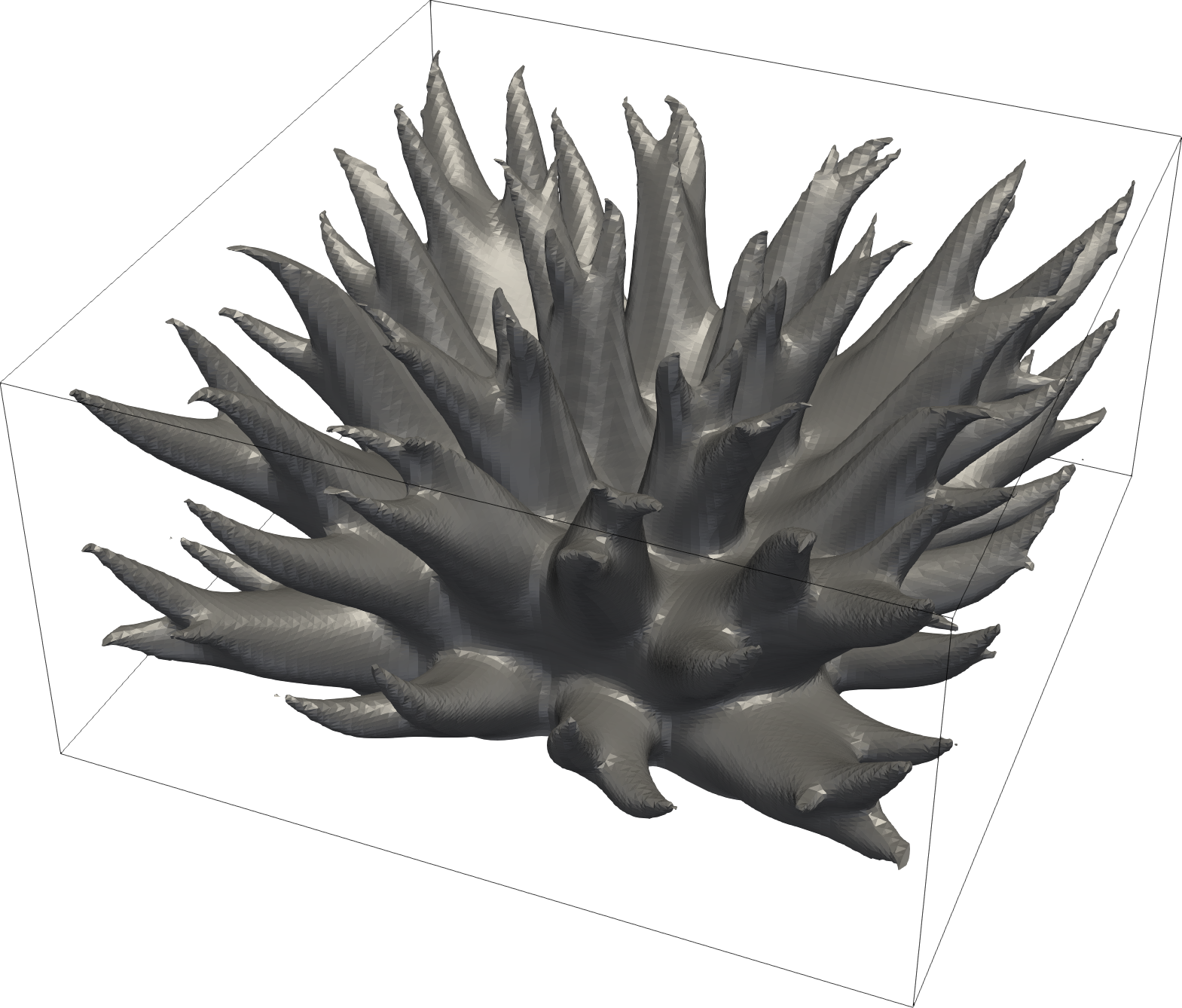}
		}\\
		\subfigure[bottom view; $\tau=1\times10^{-5}$]{
			\includegraphics[width=5cm]{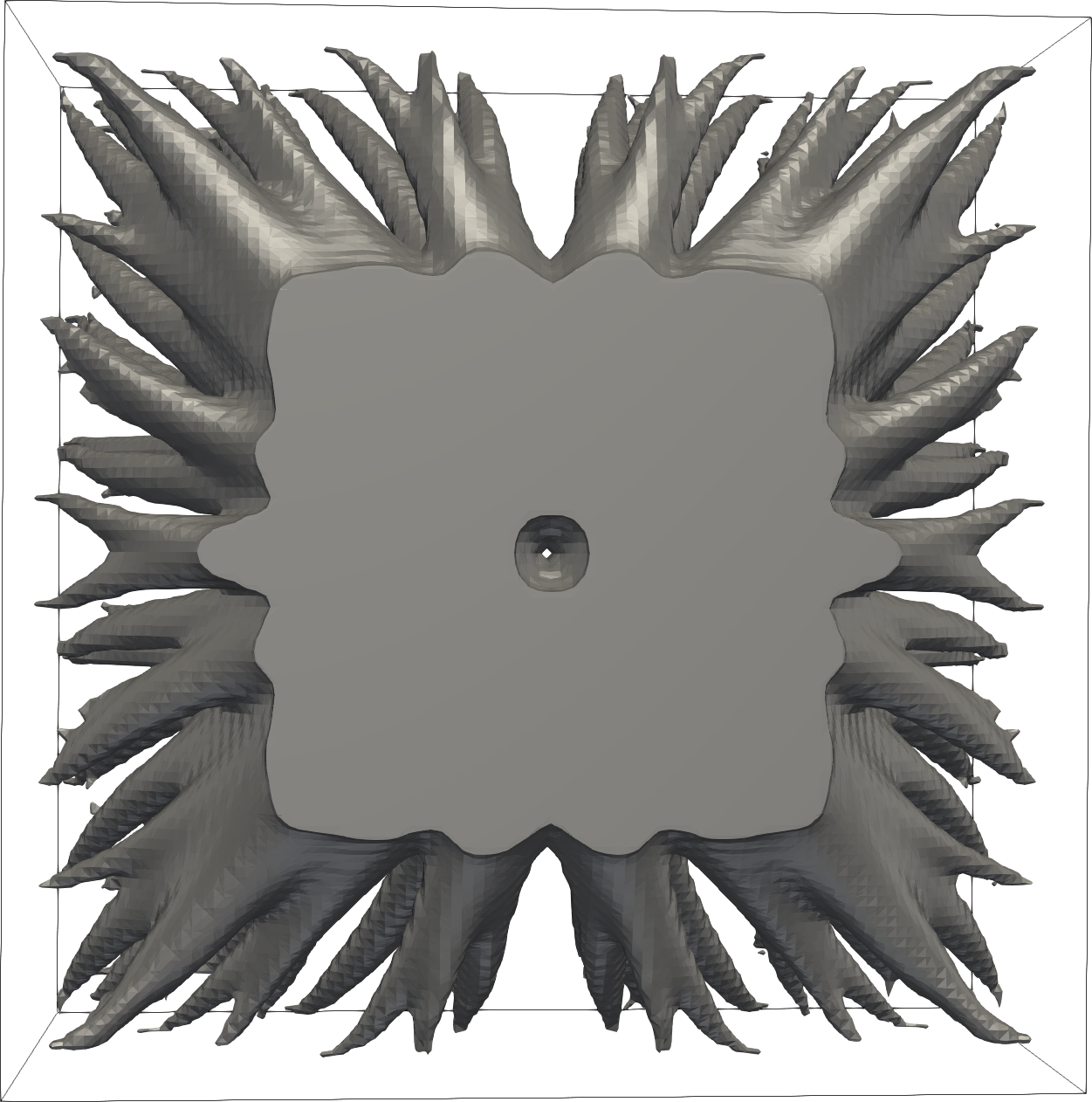}
		}\qquad \qquad
		\subfigure[bottom view; $\tau=5\times10^{-4}$]{
			\includegraphics[width=5cm]{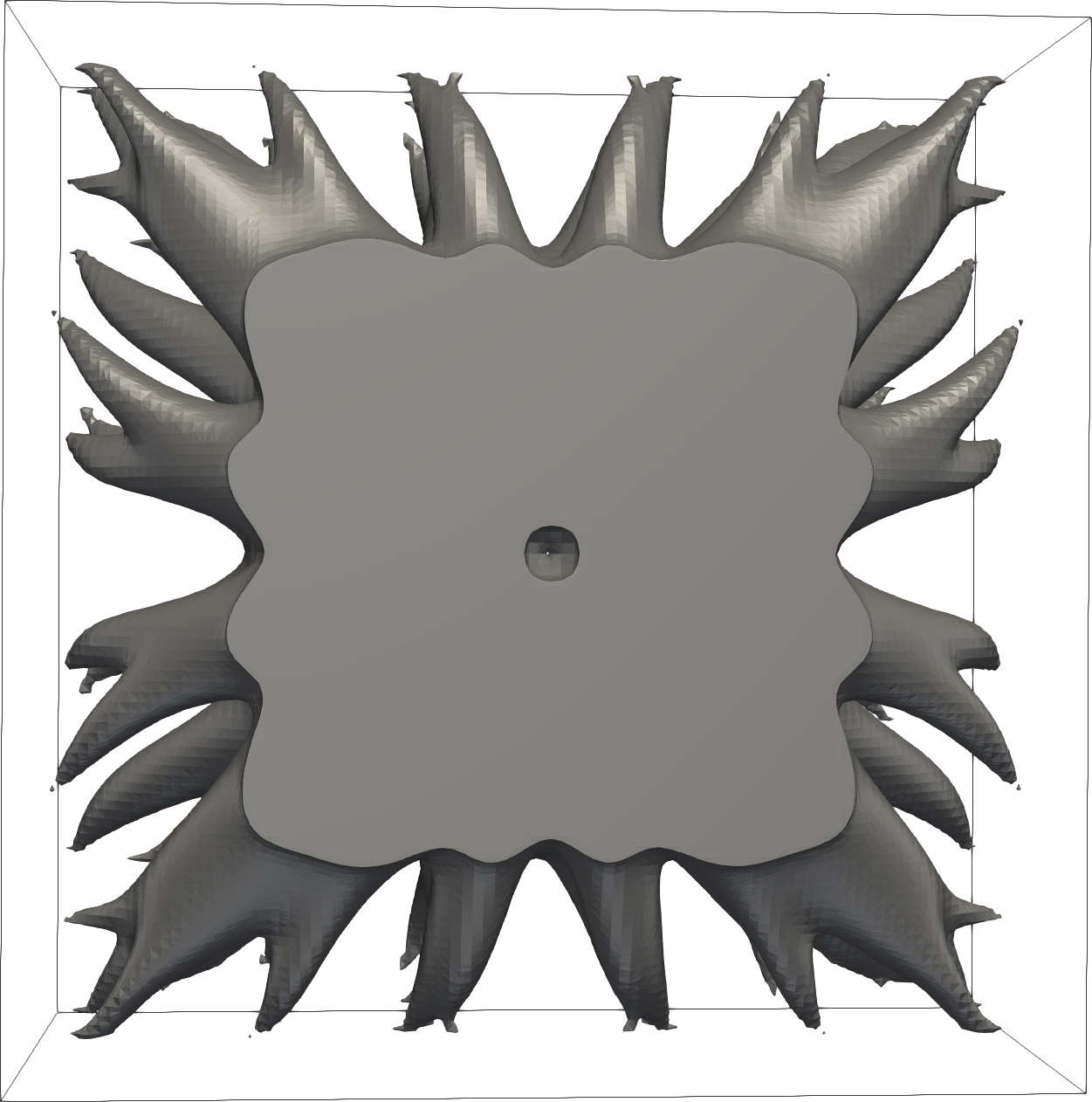}
		}
		\caption{Optimal configuration considering the geometrical constraint}
		\label{fig:thermal-with}
	\end{center}
\end{figure*}

 Here, we examine the effect that the geometric constraint has on the resulting optimal configurations. Figures \ref{fig:thermal-without} and \ref{fig:thermal-with} display the obtained optimal configurations without and with the geometric constraint (i.e., cases 1 and 2), respectively.
 \\
\\
As seen in the figures, the shapes with and without the constraint are almost identical. However, if the geometric constraint is not taken into account, an enclosed space remains in the center of the bottom surface. In contrast, if the geometric constraint is taken into account, the space in the center of the bottom surface is connected to the outside through a small hole. Thus, for the parts that do not satisfy the constraint, a solution is obtained to satisfy the desired geometric constraint with only minor shape modifications.
For this problem setup, there was no considerable difference in the shape features between the constraint-aware and constraint-unaware cases. In general, it is not possible to determine in advance whether little or no difference will be observed. For example, the case in 6.2 had considerable differences. Therefore, topology optimization with the geometric constraint should always be applied while using additive manufacturing.

\newpage
\section{Prototype example}
We fabricated the optimal shape by additive manufacturing. Naturally, manufacturing is possible only when geometric constraints are taken into account. Here, we selected a topology optimization example of the stiffness maximization problem when the regularization parameter was set to $\tau=\times 10^{-4}$. Stereolithography data were generated based on the isosurface of the level-set function using the open-source software ParaView \cite{ahrens2005paraview}.
The obtained optimal shape was printed using a LPBF machine (ProX DMP 200, 3D Systems, Inc.). The material was set to 17-4PH (LaserForm 17-4 PH, 3D Systems, Inc.).
The bedded-powder layer thickness was $30\mu \rm{m}$, hatch spacing was $50\mu \rm{m}$, the rotation angle between the bedded-powder layers was $90^\circ$, and the scanning strategy applied a hexagonal grid. The resulting shape was resized to $0.05$ times its original size.
\\
\begin{figure*}[htb]
	\begin{center}
		\includegraphics[width=13cm]{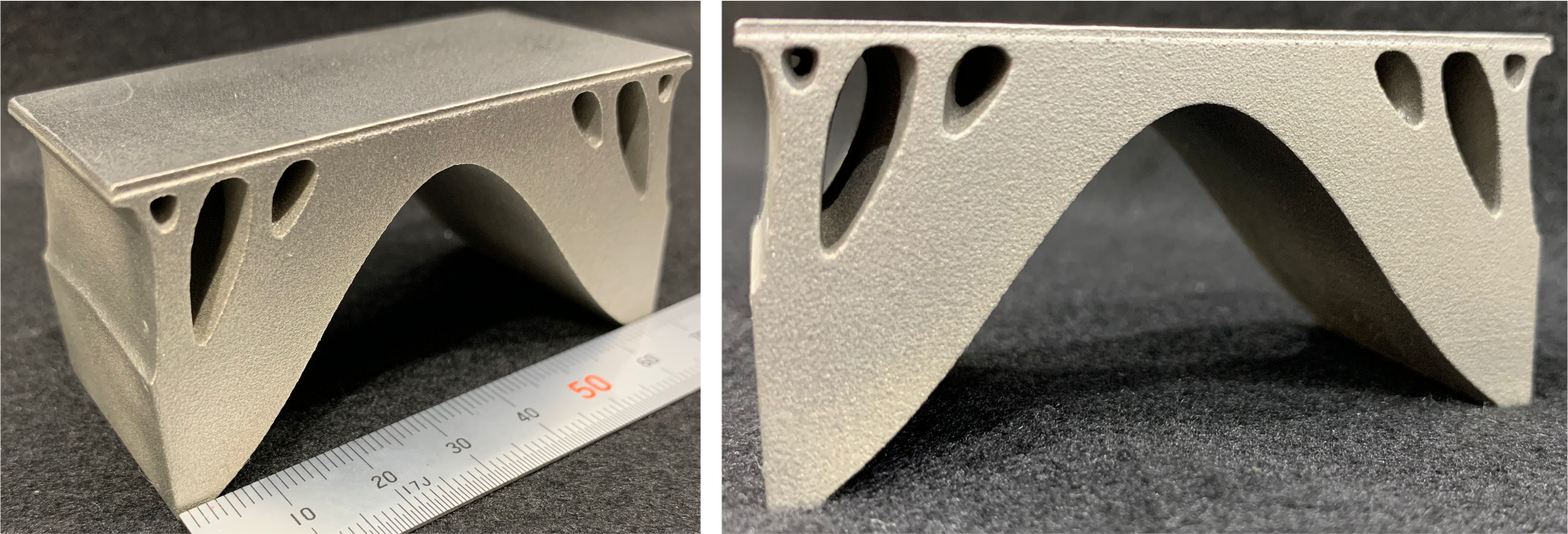}
		\caption{Prototype example}
		\label{fig:prototype}
	\end{center}
\end{figure*}\\
Figure \ref{fig:prototype} displays a photo of the manufactured prototype. The shape obtained by the proposed method could be manufactured without any shape modification.
\newpage
\section{Conclusion}
In this paper, we propose a topology optimization method that considers the no-closed-cavity constraint for additive manufacturing based on the fictitious physical model concept.
Our primary contributions are summarized as follows.
\begin{enumerate}[(1)]
	\item We discuss level-set-based topology optimization and the fictitious physical model for geometric constraints in the topology optimization method
	\item The fictitious physical model is presented for the no-closed-cavity constraint for additive manufacturing.
	\item A topology optimization problem with the geometric constraint is formulated, and an optimization algorithm is developed and incorporated into the FEM based on this formulation
		\item Numerical examples are provided to validate the proposed fictitious physical model.
	\item Optimization examples are presented that demonstrate that the proposed method can obtain optimal configurations. The proposed method makes it possible to select an appropriate structural type in consideration of the geometric constraints.
	\item A demonstration of a prototype using additive manufacturing is presented to demonstrate the effectiveness of the proposed method.
\end{enumerate}
\section*{Acknowledgments}
This work was supported in part by JSPS KAKENHI Grant Number 19H02049. The demonstration of metal additive manufacturing was assisted by Mr. Takao Miki (Osaka Research Institute of Industrial Science and Technology).

\bibliography{mybibfile}

\end{document}